\documentclass[aps,prd,twocolumn,showpacs,superscriptaddress,showkeys,nofootinbib]{revtex4} 
\usepackage{amssymb,amsmath}
\usepackage{graphicx,color} 
\usepackage{epsfig}
\usepackage{enumerate}
\usepackage{amsfonts}
\usepackage{hyperref}
\usepackage{mathrsfs}
\usepackage{latexsym}
\usepackage{natbib}
\begin{document}

\title{Chiral condensate in hadronic matter} 
\author{J.~Jankowski}
\email{jakubj@ift.uni.wroc.pl}
\affiliation{Institute for Theoretical Physics, University of Wroc{\l}aw, 
PL - 50-204 Wroc{\l}aw, Poland}
\author{D.~Blaschke}
\email{blaschke@ift.uni.wroc.pl}
\affiliation{Institute for Theoretical Physics, University of Wroc{\l}aw, 
PL - 50-204 Wroc{\l}aw, Poland}
\affiliation{Fakult\"at f\"ur Physik, Universit\"at Bielefeld, 
D - 33615 Bielefeld, Germany}
\affiliation{Bogoliubov Laboratory for Theoretical Physics, Joint Institute 
for Nuclear Research, RU - 141980 Dubna, Russia}
\author{M.~Spali\'nski}
\email{michal.spalinski@fuw.edu.pl}
\affiliation{National Center for Nuclear Research, Ho{\.z}a 69, 
PL - 00-681 Warsaw, Poland}
\affiliation{Physics Department, University of Bialystok, PL - 15-424 
Bia{\l}ystok, Poland}
\date{\today} 
\preprint{BI-TP 2013/xx}


\begin{abstract}

The finite temperature chiral condensate for $2+1$ quark flavors is
considered in the framework of the hadron resonance gas model. This requires
some dynamical information, for which two models are employed: one based on
the quark structure of hadrons combined with the Nambu-Jona-Lasinio approach
to chiral symmetry breaking, and one originating from gauge/gravity duality.
Using these insights, hadronic sigma terms are discussed in the context of
recent first principles results following from lattice QCD and chiral
perturbation theory. For the condensate, in generic agreement with lattice
data it is found that chiral symmetry restoration in the strange quark sector
takes place at higher temperatures than in the light quark sector. 
The importance of this result for a recently proposed dynamical model of 
hadronic freeze-out in heavy-ion collisions is outlined.

\end{abstract}



\maketitle



\section{Introduction}
\label{Intro}

Spontaneous chiral symmetry breaking is, apart from color confinement, the
most important physical aspect of strong interactions. The fact that one
observes mass splittings of chiral partners in the hadron spectrum and that 
pions
have properties attributed to Goldstone bosons strongly suggests that chiral
symmetry is spontaneously broken in the vacuum. These, and other theoretical
arguments \cite{Weinberg:1995mt}, imply 
that in the vacuum there exists a chiral condensate 
giving rise to an expectation value of the
bilinear fermionic operator $\bar{\psi}\psi$.  Dynamical details of this
phenomenon, which is inherently non-perturbative in nature, are part of the
long standing problem of strong interactions, but in the course of time
different model mechanisms for all its different aspects have been developed.

As temperature and/or baryon density is increased, thermal hadron excitations,
because of their quark substructure, will affect the vacuum condensate
causing its melting and eventually vanishing at the transition line to the
chirally symmetric phase. Microscopic quantification of this phenomenon comes
from first principles lattice QCD (lQCD) simulations and confirms the
intuitive predictions \cite{Borsanyi:2010bp}.

To get physical insight into this effect for low temperatures (and densities)
one can use the hadron resonance gas (HRG) model \cite{Hagedorn:1968zz}, which was previously
successfully applied to give a physical interpretation of lQCD data
\cite{Borsanyi:2010bp,Bazavov:2009zn} as well as a description of
the abundances of particles produced in heavy ion collisions
at very different center of mass energies
\cite{BraunMunzinger:2003zd,BraunMunzinger:2001ip}
in terms of freeze-out parameters. The assumption underlying this approach is that
for conditions below the QCD transition line the system is composed of
non-interacting hadronic degrees of freedom and so the partition function is
that of an ideal mixture of free quantum gases.  To have a reliable physical description of the
system one needs to take into account all hadron resonances with masses up to
$\sim2$~GeV.

To calculate the condensate in this framework it is necessary to know the
dependence of hadron masses on the current quark masses. This, apart from the
Goldstone boson octet, is not straightforward to determine and either requires
some assumptions about the underlying dynamics or is the result of a
phenomenological fit. Approaches, which can be regarded as based on first
principles, are chiral perturbation theory (ChPT) \cite{Borasoy:1996bx},
lattice QCD simulations \cite{Durr:2008zz} and Dyson-Schwinger equations (DSE)
\cite{Flambaum:2005kc,Holl:2005st}. They provide a consistent picture 
of hadrons with a reliable account of the quark
mass dependence. However, in the ChPT framework there are still large
uncertainties concerning for example the nucleon strange sigma term
\cite{Ren:2012aj} for which, when different orders of approximation are
considered, even the sign is not clear \cite{MartinCamalich:2010fp}.

This article explores the consequences of various
hadron mass formulae proposed recently and 
compares them with the results mentioned above.  

One set of mass formulae which was used in the analysis reported here
comes from a new model based on quark counting and is a generalization of what
was proposed by Leupold \cite{Leupold:2006ih} a few years ago. In Leupold's
scheme hadron masses were assumed to be linear in the current quark
masses. This approach was used (and generalized) in \cite{Blaschke:2011hm}. In
the present work a further step is taken: it is assumed that the dependence of
hadron masses on the current quark mass arises solely due to the dependence of
constituent masses of valence quarks. The response of these 
constituent masses 
to the change in the current quark mass is
determined based on the Nambu-Jona-Lasinio (NJL) model \cite{Klevansky:1992qe,Hatsuda:1994pi,Rehberg:1995kh}.
In this way a fairly good description of the hadronic sigma terms is
obtained. The only flaw is that the sea-quark contributions are neglected
entirely, which, 
for example, leads to the vanishing of the nucleon strange sigma term.

The second approach considered in this paper is the use of baryon mass
formulae which were obtained in a 
large $N$ \cite{'tHooft:1973jz} holographic model of QCD due to Sakai and
Sugimoto \cite{Sakai:2004cn}. The last ten years have witnessed a lot of
progress coming from gauge/gravity duality allowing for valuable insights into
the dynamics of strongly coupled gauge theories. Recent developments have made
it possible to study in a quasi-analytical way theories which have very
realistic properties. The spectrum of mesons and chiral symmetry breaking in
the chiral limit was studied in \cite{Sakai:2004cn} and
static baryon properties (such as masses, magnetic moments or charge radii)
\cite{Hata:2007mb,Hashimoto:2008zw} were found to be in  qualitative
agreement with experiment. Also form factors
\cite{Hashimoto:2008zw} agree quite well with the data. Further progress was
made with the extension to finite current quark masses \cite{Hashimoto:2008sr}
(see \cite{Bergman:2007pm} for an alternative construction) where for example
Gell-Mann-Oakes-Renner relations for the pseudoscalar octet have been
demonstrated. The impact of finite current quark masses on the spectrum of nucleon
octet and delta decuplet baryons has been considered in the two flavor case
\cite{Hashimoto:2009hj} and for $2+1$ flavors \cite{Hashimoto:2009st} with
nontrivial results. The leading order corrections are proportional to the
squares of Goldstone boson masses and determined in a similar way as the
leading order of ChPT \cite{MartinCamalich:2010fp}. The results are in good 
agreement with the data and
other theoretical expectations in the light quark sector, while the
contribution of the strange quark is overestimated by the model. It is very
likely that going beyond the leading order in the expansion in powers of the
current quark masses will give more reasonable results (as is the case in
ChPT). Also, at leading order, vector mesons were argued not to receive mass
corrections from finite current quark mass \cite{Hashimoto:2009hj}. 
The mass formulae obtained in \cite{Hashimoto:2009hj,Hashimoto:2009st} make it
possible to estimate sigma terms, including those for the nucleon octet and
delta 
decuplet. This is then used to calculate the chiral condensate in the
framework of the HRG model and the results are compared with calculations based
on chiral perturbation theory.

In the context of DSE studies \cite{Flambaum:2005kc,Holl:2005st}
sigma terms for the two light quark flavours have been considered. 
In addition to the nucleon and delta baryons also vector mesons were
included. Due to the $\rho-\pi\pi$ and $\rho-\pi-\omega$ couplings 
one gets a sigma term of the $\rho$-meson.  
The $\omega$-meson has no pion loop dressing and
therefore only $\omega-\rho\pi$ coupling remains. 
Since in the DSE approach strange sigma terms were not included yet we do not 
use these results as a base for calculating the chiral condensates of 
interest. We will only use it as a reference point to other calculations. 

The importance of hadronic contribution to the melting of the chiral
condensate was appreciated in a model for the freeze-out stage of heavy-ion
collisions, where it was related to the Mott-Anderson delocalisation of
hadrons \cite{Blaschke:2011hm}. 
The model is based on assumptions for hadron-hadron interactions and on the 
evolution of the matter formed in heavy ion collisions.
The main point is that freeze-out phenomena are
assumed to take place in the hadronic phase and are entirely attributed to the
hadron dynamics. In general, each hadron is assigned a medium dependent radius
$r_h(T,\mu_B)$, which is then related in a universal way to the chiral
condensate. Hadron-hadron reactions are described by the Povh-H\"ufner
law \cite{Hufner:1992cu} and in consequence the cross section is determined by
the medium dependence of the condensate. 
As the temperature decreases, the mean time between the interactions is 
getting larger, since it is inversely proportional to the reaction 
cross-section and hadron density (in the {\it relaxation time approximation}). 
At some point the reaction rate becomes smaller than the rate of expansion 
and reactions between hadrons stop to change the final composition. 
The freeze-out parameters $\mu_B^f$ and $T^f$ are determined by the equality 
of both time scales. 
However in Ref.~\cite{Blaschke:2011hm} only the light quark condensate was 
considered, so one of the possible improvements of the model is to include 
also the strange sector. 
This is one of the motivations for the present studies.

The organization of the paper is as follows: in section \ref{HRGM} the generic
theoretical setup is described and the relevant quantities used in further
calculations are defined. This section also reviews some thermodynamic
quantities computed in the HRG model and highlights very good agreement with
lattice computations. 
These considerations do not require any detailed assumptions about hadron 
dynamics. 
On the other hand, the computation of the chiral condensate strongly depends 
on hadron mass formulae expressed in terms of current quark masses as 
discussed in section \ref{Condensate}. 
This dependence is captured by the hadronic sigma terms. 
In section \ref{ChPT} we describe our baseline which are results obtained 
within ChPT as the low energy effective theory of QCD.  
In section \ref{QWC} hadron mass formulae based on their constituent quark 
structure are presented and contrasted with a previously established 
parametric dependence 
\cite{Karsch:2003vd,Karsch:2003zq} and with first principles results. 
Section \ref{Holo} contains novel results following from the Sakai-Sugimoto 
holographic model together with a discussion in the light of lowest order ChPT 
results. 
Section \ref{Conclusions} contains conclusions, some discussion and possible 
open directions.


\section{Hadron Resonance Gas Model}
\label{HRGM}

The hadron resonance gas model implements the idea
\cite{Hagedorn:1968zz} 
that QCD thermodynamics in the hadronic phase 
can be described as a multicomponent ideal hadron gas. 
For very low temperatures the dominant degrees of
freedom are pions and kaons and due to the Goldstone theorem
their interactions are weak. Therefore in the first approximation they 
can be considered as free particles. As the temperature and/or 
density is increased, contributions from heavier hadrons
become important. Because strong interactions are of 
finite range in the thermodynamical limit of infinite
volume $V\rightarrow\infty$ the grand canonical 
potential\footnote{Since only homogeneous systems are considered here, the
  symbol $\Omega$ denotes the grand canonical potential 
  as usually defined in statistical mechanics
  divided by the volume.} 
can be expressed as a sum of contributions from free
hadrons \cite{Dashen:1969ep}
\begin{equation}
\Omega(T,\{\mu_i\}) = \Omega_0 + \Omega_{\rm HRG}(T,\{\mu_i\})~. 
\label{eq:Omega} 
\end{equation}
In the above formula
$\Omega_0$ is the vacuum part, whose
detailed form is irrelevant for the following considerations, 
and the medium dependent part contains 
contributions from mesons and baryons 
\begin{equation}
\Omega_{\rm HRG}(T,\{\mu_i\}) = \Omega_{\rm M}(T, \{\mu_i\}) + \Omega_{\rm B}(T,\{\mu_i\})~.
\label{eq:}
\end{equation} 
Here $\{\mu_i\}$ is the set of chemical potentials corresponding to conserved
charges such as baryon number $B$, electric charge $Q$, isospin $I_3$
and strangeness $S$. The free meson contribution reads 
\begin{equation}
\Omega_{\rm M}(T,\{\mu_i\}) = \sum_M d_M \int \frac{d^3k}{(2\pi)^3}T\ln(1-z_Me^{-\beta E_M})~, 
\label{eq:OmegaMeson}
\end{equation}
while the free baryon contribution is 
\begin{equation}
\Omega_{\rm B}(T,\{\mu_i\}) = -\sum_B d_B \int \frac{d^3k}{(2\pi)^3}T
\ln(1+z_Be^{-\beta E_B})~, 
\label{eq:OmegaBaryon}
\end{equation}
where $E_i=\sqrt{k^2+m_i^2}$
and $d_B$ and $d_M$ count the 
degeneracy of hadrons. 
Fugacities are defined by 
\begin{equation}
z_j = \exp\left(\beta \sum_a X^a\mu_{x^a}\right)~,
\label{eq:}
\end{equation}
where the index $a$ runs over all conserved charges in 
the system, $X^a$ is the corresponding charge and $\beta=1/T$
is the inverse temperature. Although inclusion of
chemical potentials is straightforward in the HRG approach, 
for much of this paper they are all set to zero.
The residual repulsive interactions
can be taken into account, e.g., by the van der Waals excluded
volume corrections \cite{Rischke:1991ke}.

All thermodynamic quantities, 
such as equations of state 
for pressure and energy density
as well as material properties 
such as the speed of sound, can be obtained
from the grand canonical 
thermodynamical potential $\Omega(T,\{\mu_i\})$.
In the following, let us discuss more in detail the case of vanishing 
chemical potentials.
The pressure
is given by
\begin{equation}
p = - \Omega(T,\left\{\mu_i=0\right\})~,
\label{eq:}
\end{equation}
and the energy density is 
\begin{equation}
\varepsilon =  
\frac{\partial[\beta\Omega(T,\left\{\mu_i=0\right\})]}{\partial\beta}~. 
\label{eq:}
\end{equation}
%


\begin{figure}[!htb]
\includegraphics[width=.45\textwidth]{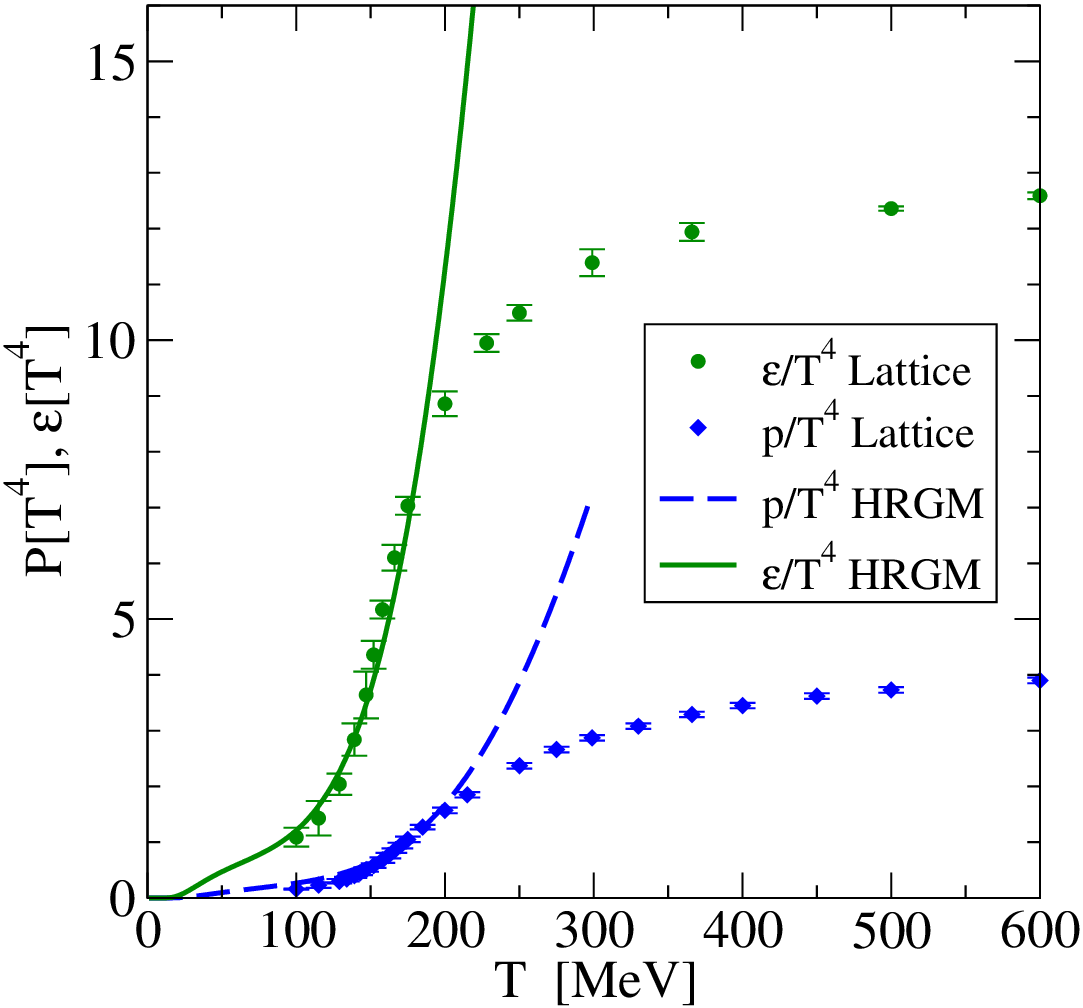}
\caption{ (Color online). Energy density and pressure for the HRG compared to
lQCD data  \cite{Borsanyi:2010cj}. The upper limit for the mass of hadrons
included in the calculation is $m_{\rm max}=2$~GeV.}%
\label{PEplot}%
\end{figure}


Fig.~\ref{PEplot} 
shows the equations of state as obtained for the HRG and
compares it with recent lQCD simulations \cite{Borsanyi:2010cj}
normalized to $T^4$, the Stefan-Boltzmann behaviour of a massless ideal gas. 
There is clearly an excellent agreement for
temperatures up to $\sim 170$~MeV which means
that the dominant effect in that range of
temperatures comes from the excitation of hadronic
degrees of freedom rather than from their interactions.
This is a very well known effect \cite{Borsanyi:2010bp,Bazavov:2009zn}.
Agreement for higher temperatures can 
be obtained when medium modifications
of hadronic states are taken into account.
For example in \cite{Turko:2011gw} it was demonstrated that inclusion of 
state dependent hadronic width $\Gamma_h(T)$ taken on the inverse
collision time of the Mott-Anderson freeze-out \cite{Blaschke:2011hm} 
and proper introduction of quark-gluon degrees of freedom
based on the Polyakov-loop extended Nambu-Jona-Lasinio (PNJL) model
nicely reproduces lattice QCD data in the whole temperature range.

\begin{figure}[!htb]
\includegraphics[width=.45\textwidth]{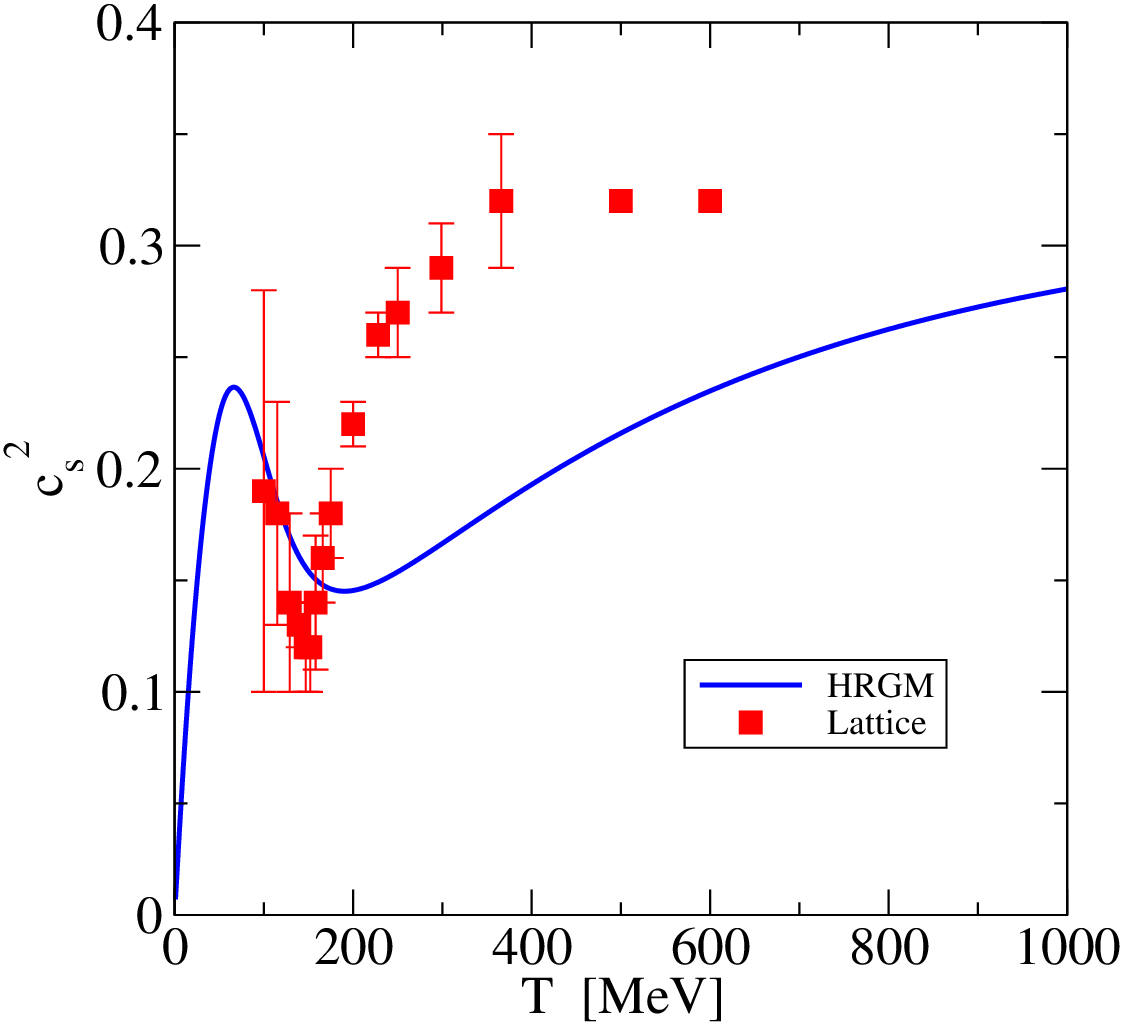}
\caption{ (Color online). Squared sound velocity for HRG compared to
lQCD data  \cite{Borsanyi:2010cj}. The upper limit for the mass of hadrons
included in the calculation is $m_{\rm max}=2$~GeV.}%
\label{cs}%
\end{figure}

The velocity of sound is given by
\begin{equation}
c_s^2 = \frac{dp}{d\varepsilon} = 
\frac{\varepsilon+p}{T}\left(\frac{d\varepsilon}{dT}\right)^{-1}~,
\label{eq:}
\end{equation}
where the second equality holds only for zero chemical potentials.
Its temperature dependence is shown in Fig.~\ref{cs} for the HRG
model compared to lQCD data \cite{Borsanyi:2010cj}. 


Qualitatively
in the HRG model the increase for low temperatures
is related to the appearance of a large number of light 
degrees of freedom. When heavier hadrons are exited, 
they contribute considerably to the energy density but almost 
nothing to the pressure, which leads to the characteristic dip. 
For high temperatures, because the number of states 
included is finite, there is an approximately constant behavior approaching
the massless gas limit $c_s^2=1/3$ only for very high temperatures.
On the other hand, for lQCD the dip is an indicator 
of the crossover transition.
For a first order transition, the sound velocity should be strictly zero.
For high temperatures lattice data approach the massless limit, which is 
consistent with the interpretation 
of deconfinement to a massless quark-gluon
medium.

The importance of the speed of sound
for the phenomenology of heavy ion collisions was noticed, e.g.,  
by Florkowski {\it et al.} 
\cite{Ryblewski:2010tn,Florkowski:2008es}  
in the context of the HBT puzzle.


\section{Chiral condensate and sigma terms}
\label{Condensate}

Using the standard formula for the chiral condensate
\begin{equation}
\langle\bar{q}q\rangle= \frac{\partial\Omega(T,\{\mu_i\})}{\partial m_0}~,
\label{eq:LightCondenste}
\end{equation}
one obtains the quark-antiquark condensate in the light 
and strange flavor  sector respectively
\begin{equation}
\langle\bar{q}q\rangle= \langle\bar{q}q\rangle_0 + \frac{\partial\Omega_{HRG}(T,\{\mu_i\})}{\partial m_q}~,
\label{eq:LightCondenste}
\end{equation}
\begin{equation}
\langle\bar{s}s\rangle = \langle\bar{s}s\rangle_0 + \frac{\partial\Omega_{HRG}(T,\{\mu_i\})}{\partial m_s}~.
\label{eq:sCondenste}
\end{equation}
%
The derivatives are taken with respect to the current
quark masses and lead to the generic formulae
\begin{eqnarray}
\langle\bar{q}q\rangle & = & \langle\bar{q}q\rangle_0 + \sum_M\frac{\sigma_q^M}{m_q}n_M(T,\{\mu_i\})\\
\nonumber
&+& \sum_B\frac{\sigma_q^B}{m_q}n_B(T,\{\mu_i\})~,
\label{eq:LightCondensate}
\end{eqnarray}
\begin{eqnarray}
\langle\bar{s}s\rangle & = & \langle\bar{s}s\rangle_0 + \sum_M\frac{\sigma_s^M}{m_s}n_M(T,\{\mu_i\})\\
\nonumber
&+& \sum_B\frac{\sigma_s^B}{m_s}n_B(T,\{\mu_i\})~,
\label{eq:sCondensate}
\end{eqnarray}
where the scalar densities of mesons and baryons
have been introduced as
\begin{equation}
n_M(T,\{\mu_i\}) = \frac{d_M}{2\pi^2} \int_0^\infty~dk k^2 \frac{m_M}{E_M} 
\frac{1}{z_M^{-1}e^{\beta E_M}-1}~,
\label{eq:}
\end{equation}
\begin{equation}
n_B(T,\{\mu_i\}) = \frac{d_B}{2\pi^2} \int_0^\infty~dk k^2 \frac{m_B}{E_B} 
\frac{1}{z_B^{-1}e^{\beta E_B}+1}~,
\label{eq:}
\end{equation}
and the response of hadron masses to changes in the current 
quark mass of flavor $f=u,d,s,...,q_{N_f}$ is captured by the 
hadron sigma terms 
\begin{equation}
\sigma_f^h = m_f \frac{\partial m_h}{\partial m_f}~.
\label{eq:}
\end{equation}
Thus, for every hadron state,
there are different sigma terms related to contributions from 
quark flavors constituting the hadron. 

The above formulas are valid for the non-interacting gas.
In the light flavour sector effects of meson-meson 
and pion-nucleon interactions 
as described by ChPT were implemented in \cite{GarciaMartin:2006aj}.

In the following, isospin symmetry is assumed, setting the light quark masses
$m_q=m_u=m_d\approx 5.5$~MeV 
and the light quark condensate
$\langle\bar{u}u\rangle_0=\langle\bar{d}d\rangle_0=\langle\bar{q}q\rangle_0=(-240)^3~\textrm{MeV}^3$. 
Analysis based on lQCD and QCD sum rules together with the low energy theorem
for the correlation functions allows one to estimate the ratio of 
strange to light quark condensates to be $0.8\pm0.3$
\cite{Jamin:2002ev} (other estimates give $0.75\pm0.12$
\cite{Narison:2002hk} but note also recent explicit
lattice calculations \cite{McNeile:2012xh}). 
One can understand this hierarchy of condensates
using the spectral representation of the expectation value
for the quark of current mass $m_f$ \cite{Banks:1979yr,Langfeld:2003ye}
\begin{equation}
\langle\bar{q_f}q_f\rangle = -2m_f \int_0^\infty d\lambda 
\frac{\rho(\lambda)}{\lambda^2+m_f^2}~,
\label{eq:SpectralRepresentation}
\end{equation}
and noting that the spectral integral is increasingly suppressed with the 
higher current quark mass, thus lowering the value of the quark condensate. 
Furthermore, the characteristic length scale
related to the quark-antiquark condensate can be taken as
$1/m_f$ which is smaller for greater masses. This implies
that the medium effect -- expressed as screening length -- will
affect heavier quark condensates at higher temperatures.
This can also be understood as arising from the fact that the contribution
to the strange quark condensate -- and its melting --  comes
from strange hadrons, which are fewer in number than hadrons 
containing light quarks. 

Another important quantity, an approximate order parameter 
for the deconfinement phase transition, is the Polyakov loop.
It is very well studied in lQCD and 
recently it has been addressed within the HRG framework
\cite{Megias:2012kb}.
Good agreement with the lattice data was found in the 
temperature range $150$~MeV$<T<190$~MeV.


\section{Hadron masses in chiral perturbation theory}
\label{ChPT}

As explained above, the finite temperature behavior of chiral condensates in
the HRG approach is determined by the sigma terms, which express the
dependence of hadron masses on the current quark masses. A very important
approach to this problem is provided by chiral perturbation theory. This
approach is most effective in the pseudoscalar sector, since in the limit of
vanishing quark masses these states are massless Goldstone bosons of
spontaneously broken chiral symmetry. The importance of the chiral
perturbation theory results for the sequel is twofold. Firstly, in the following
section they are used to compute sigma terms for the pseudo-Goldstone bosons
-- the model introduced there is used for the remaining hadronic
states. Secondly, it is a natural point of reference for calculations carried
out in section \ref{Holo}, where a detailed comparison with the holographic
approach is described.

In the case of the Goldstone boson octet the relevant mass formula is the
Gell-Mann-Oakes-Renner (GMOR) relation, which takes the form
\cite{Gasser:1984gg} 
\begin{equation}
f_\pi^2m_\pi^2\left(1-\kappa \frac{m_\pi^2}{f_\pi^2}\right) = 
- \langle\bar{q}q\rangle_0(m_u+m_d)~,
\label{eq:GMORpion}
\end{equation}
\begin{equation}
f_K^2m_K^2\left(1-\kappa \frac{m_K^2}{f_\pi^2}\right) = 
- \frac{\langle\bar{q}q\rangle_0 + \langle\bar{s}s\rangle_0}{2}(m_q+m_s)~.
\label{eq:GMORkaon}
\end{equation}
These formulae include next to leading order corrections expressed in terms of
the parameter $\kappa=0.021\pm0.008$  \cite{Jamin:2002ev}.
If one assumes $\langle\bar{s}s\rangle_0=0.8\langle\bar{q}q\rangle_0$,~  
$f_\pi=92.4$~MeV, $f_K=113$~MeV
(which gives $f_K/f_\pi\approx1.22$ \cite{Leutwyler:1984je})
and $m_s=138$~MeV, then one finds 
$(m_q+m_s)/m_s\approx 1.040$
as compared to the lattice choice \cite{Borsanyi:2010bp}
$(m_q+m_s)/m_s\approx 1.036$.

Taking the derivative of the above equations with
respect to the light quark masses one finds 
\begin{equation}
\frac{\partial m_\pi^2}{\partial m_q} = 
- \frac{\langle\bar{q}q\rangle_0}{f_\pi^2\left(1-2\kappa \frac{m_\pi^2}{f_\pi^2}\right)}
\approx
- \frac{\langle\bar{q}q\rangle_0}{f_\pi^2}\left(1+2\kappa \frac{m_\pi^2}{f_\pi^2}\right)~,
\label{eq:PionDer}
\end{equation}
and similarly for the derivatives of the kaon mass with respect to $m_q$ 
(and $m_s$):
\begin{eqnarray}
\frac{\partial m_K^2}{\partial m_{q,s}} & = &
- \frac{\langle\bar{q}q\rangle_0+\langle\bar{s}s\rangle_0}{2f_K^2\left(1-2\kappa\frac{m_K^2}{f_\pi^2}
\right)}\\ 
\nonumber
&\approx&
- \frac{\langle\bar{q}q\rangle_0+\langle\bar{s}s\rangle_0}{2f_K^2}\left(1+2\kappa\frac{m_K^2}{f_\pi^2}
\right)~.
\label{eq:KaonDer}
\end{eqnarray}

In ChPT mass formulae for the ground state baryons  
can be also computed in the $N_f=2$ \cite{Bernard:2003rp} 
and $N_f=2+1$ cases \cite{MartinCamalich:2010fp}.
At lowest order 
the shift due to the finite current quark mass is proportional to the square of the Goldstone
boson mass \cite{MartinCamalich:2010fp}. 
The lowest order contributions to the baryon masses read
\begin{equation}
M_B = M_B^ 0 - \sum_{\phi=\pi,K}\xi_{B,\phi}m_\phi^2~, 
\label{eq:}
\end{equation}
where $\xi_{B,\phi}$ are expressed in terms of the parameters
of the low energy Lagrangian. Sigma terms
following from this formula are listed in the table
\ref{tab:sigmaChPT}. In section
\ref{Holo} these results will compared with the holographic mass formulae. 


Higher order contributions are given by the loop corrections
to baryon self energies and are evaluated using different regularization 
schemes. All of them are carefully compared to the recent lattice data.
It is interesting to note that the strange nucleon sigma term turns 
out to be negative at next-to-leading order (NLO) 
$\sigma^N_s\approx-4$~MeV \cite{MartinCamalich:2010fp}
which means that the nucleon mass decreases if the strange quark mass is
raised.  
On the other hand different ChPT studies give
at N$^3$LO 
$\sigma^N_s\approx130$~MeV\cite{Ren:2012aj}
which means that higher order corrections can be 
relevant and the existing answers must be regarded as somewhat tentative. 
First principle lattice simulations with $N_f=2+1$ dynamical
quark flavours give $\sigma^N_s=49\pm25$~MeV \cite{Junnarkar:2013ac}.


\begin{widetext}
\begin{center}
\begin{table}[h!]
		\begin{tabular}{c|*{8}{c}}
           & $N$     &  $\Lambda$ &  $\Sigma$ &   $\Xi$ &  $\Delta$ & $\Sigma^*$  & $\Xi^*$  & $\Omega^-$      \\ \hline
			 $\sigma_q$    & 36.2527 & 19.8647 & 10.7535 & 27.8369 & -1.07895 & 15.5728  & 3.30878 & -8.95528 \\ 
			 $\sigma_s$    & 162.474 & 437.693 & 590.705 & 317.096 & 789.418 & 523.058 & 729.019 & 934.981
		\end{tabular}
		\caption{Sigma terms for the lowest lying baryons in the leading order ChPT in MeV}
		\label{tab:sigmaChPT}
\end{table}
\end{center}
\end{widetext}


In this context an interesting quantity to look at 
is strangeness content of the nucleon which was estimated in
the form \cite{Borasoy:1996bx}
\begin{eqnarray}
\label{yEqnzzz}
y &=& \frac{2\langle p|\bar{s}s|p\rangle}{\langle p|\bar{u}u+\bar{d}d|p\rangle} \\
\nonumber
&=& \frac{m_\pi^2}{\sigma_{\pi N}}\left(m_K^2-\frac{1}{2}m_\pi^2\right)^{-1}m_s\frac{\partial m_N}{\partial m_s}\approx 0.21~, 
\end{eqnarray}
where $|p\rangle$ is a nucleon state of momentum $p$. 
This is similar to the famous Wroblewski 
factor \cite{Wroblewski:1985sz} introduced in
heavy-ion and pp collisions for quantifying 
strangeness production. For the $SU(3)$ symmetric
case $y=1$ while when there are no strange 
quark pairs $y=0$. The second equality of Eq. (\ref{yEqnzzz})
is quite generic and relies only on the Hellman-Feynman
theorem and tree level GMOR relations.
Its importance lies in the fact that 
making some statement about the nucleon strange
sigma term is in fact equivalent to
making a statement about the strangeness contribution
to the nucleon.


\section{Constituent quark picture}
\label{QWC}

The model described in this section is 
based on the valence quark structure of hadrons and is a nontrivial
generalization of the formulae used in \cite{Blaschke:2011hm} for the light
quark condensate (following earlier work by Leupold \cite{Leupold:2006ih}). 
This model is also compared with a another approach which gives 
a parametric dependence of hadron 
masses on the pion mass \cite{Karsch:2003vd,Karsch:2003zq} and was previously used for the
calculation of both light and strange quark condensates \cite{Tawfik:2005qh}.
In these two models mass formulae are in principle given for all the hadron
states and so the sums over mesons ($M$) and baryons ($B$) which appear in the
HRG model take into
account all states up to mass $\sim2$~GeV.

The scenario introduced here assumes  
that baryon and meson masses scale as 
\begin{equation}
m_B = (3-N_s)M_q + N_sM_s+\kappa_B~,
\label{eq:L1}
\end{equation} 
\begin{equation}
m_M = (2-N_s)M_q + N_sM_s + \kappa_M~.
\label{eq:L2}
\end{equation}
Equation (\ref{eq:L1}) is used for all baryonic 
states while equation (\ref{eq:L2}) is used for 
all mesons except pions and kaons, for which
the GMOR relations of the previous section are employed.
The quark masses in mass formulae (\ref{eq:L1}) and (\ref{eq:L2})
are the dynamical (constituent) ones and are denoted by $M_q$ 
for the light quarks and by $M_s$ for
the strange quark. 
The parameter $N_s$ measures the strangeness content
of the hadron and the quantities $\kappa_B$, $\kappa_M$ depend on the state,
but not on the current quark masses.
For the open strange hadrons $N_s$ is simply 
the number of strange (anti-strange) quarks.
For hidden strange mesons 
-- such as for example the  $\eta$ or $h_1$ state -- 
it is modified  
by the squared modulus of the coefficient
of the $\bar{s}s$ contribution to the meson wave function.
There are two possible wave function assignments 
related to the flavor singlet
$\psi_0=\frac{1}{\sqrt{3}}(\bar{u}u+\bar{d}d+\bar{s}s)$
and flavor octet 
$\psi_8=\frac{1}{\sqrt{6}}(\bar{u}u+\bar{d}d-2\bar{s}s)$
wave functions
for the hidden strange mesons.
The strangeness counting parameters  
$N_s^{(0)}=2/3$ for the singlet and $N_s^{(8)}=4/3$
for the octet have been adopted. 

It is easy to see that
the baryon octet Gell-Mann-Okubo relation:
$3M_\Lambda+M_\Sigma=2(M_N+M_\Xi)$
is translated into a constraint on the state
dependent contributions:
$3\kappa_\Lambda+\kappa_\Sigma=2(\kappa_N+\kappa_\Xi)$.

Two further simplifying assumptions
are made: excited states are assumed to have the same
flavour structure of the
wave functions as their respective ground states, and any possible mixing between
octet and singlet states 
(such as $\eta-\eta'$ mixing) is neglected.

\begin{figure}[!b]
\includegraphics[width=0.45\textwidth]{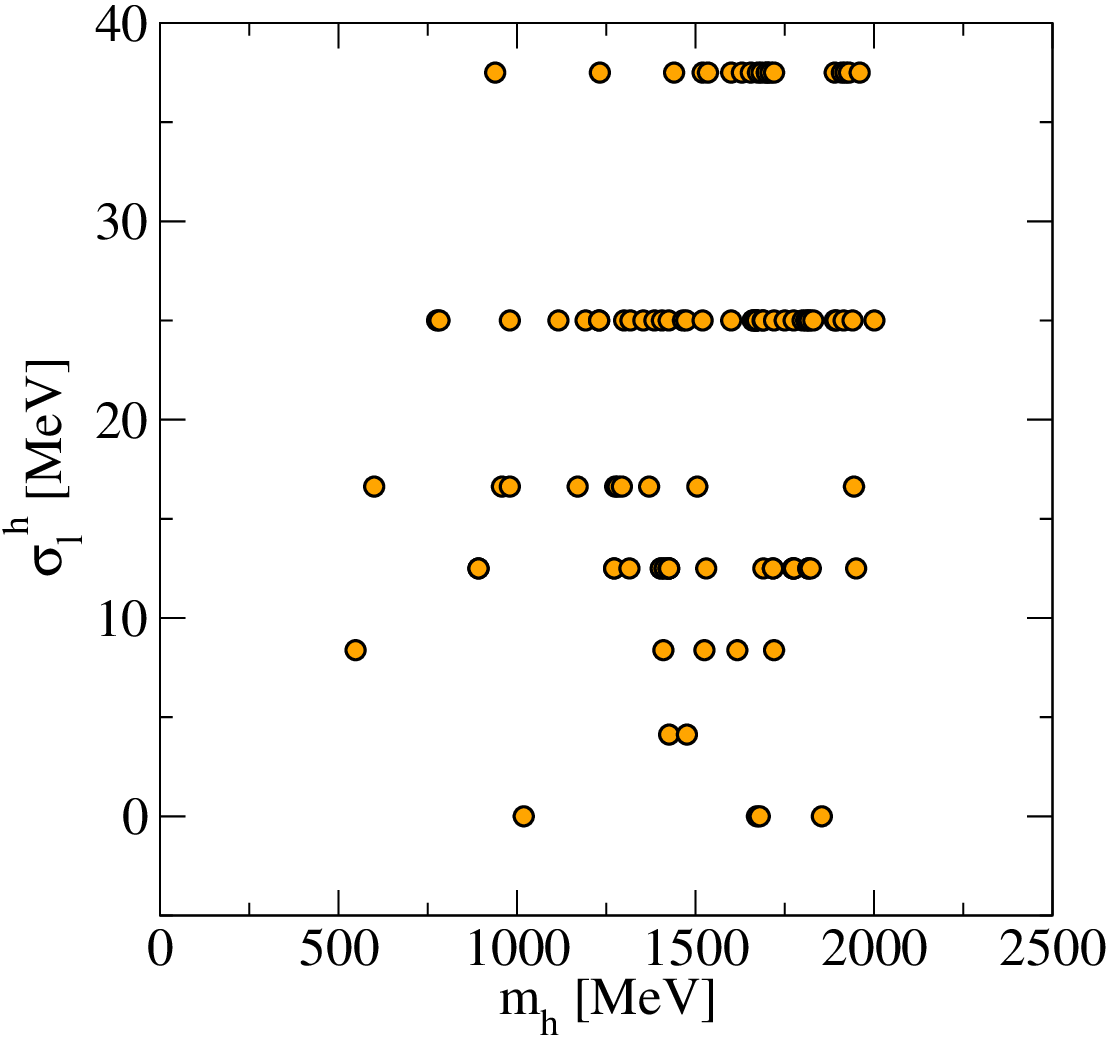}
\caption{ (Color online). Light sigma terms calculated with the constituent quark picture.}
\label{fig:SigmaL}
\end{figure}

\begin{figure}[!htb]
\includegraphics[width=0.45\textwidth]{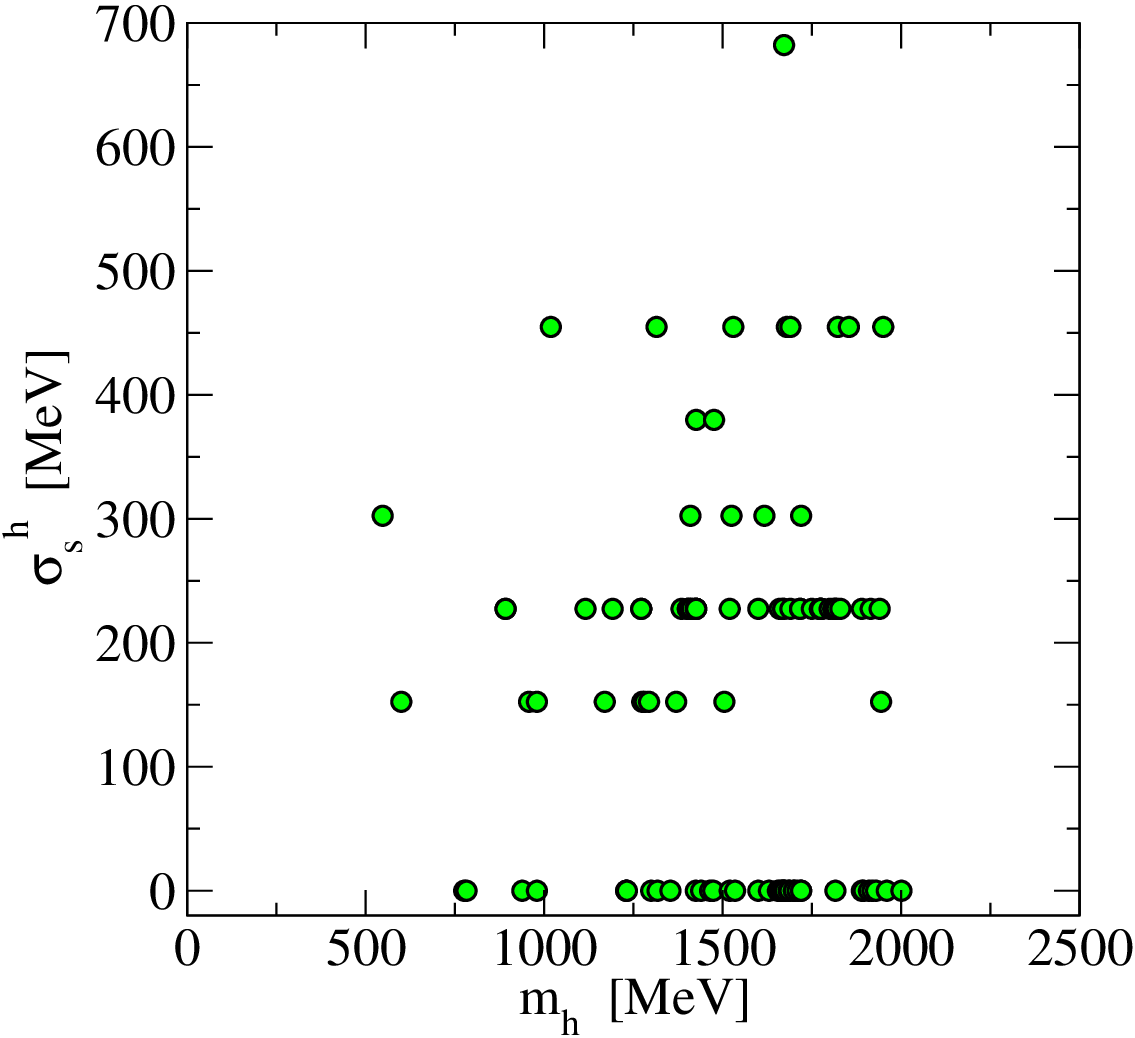}
\caption{ (Color online). Strange sigma terms calculated with the constituent 
quark picture.}
\label{fig:SigmaS}
\end{figure}

The dynamical (constituent) quark masses $M_q$ and $M_s$ appearing in 
Eqs. (\ref{eq:L1}), (\ref{eq:L2}) 
are a way of partially accounting for the dynamics of strong
interactions. For the purposes of computing the condensates only the
dependence of these constituent masses on the current quark masses is
relevant. 
This dependence is taken from the NJL model, where
the dynamically generated mass changes by 
$\Delta M_q = 12.5$~MeV as the quark mass is turned on from zero in  
the chiral limit to $m_q=5.5$~MeV ~\cite{Wergieluk:2012gd}.
This gives the nucleon sigma term
$\sigma_N=37.5$~MeV. For the strange quark
mass the value of the dynamical quark mass
is $M_s=587.4$~MeV for $m_s=140.7$~MeV which 
gives $\Delta M_s=227.4$~MeV \cite{Blaschke:2011yv}.
This valence quark counting implies that
the strange contribution to the nucleon is zero 
which is an approximation hard to control. 
For the $\Lambda$ baryon which has one strange 
quark the same arguments as above lead to the estimate 
$\sigma^\Lambda_s=252.9$~MeV.
The resulting sigma terms are shown in Figs.~\ref{fig:SigmaL},\ref{fig:SigmaS}.


In principle the scaling of Eqs.~(\ref{eq:L1}), (\ref{eq:L2}) will be 
corrected by various effects such as contributions of the sea quarks, which 
one would expect to give a logarithmic correction $\ln(m_q/\Lambda_{QCD})^2$
at one loop order. 

At this point it is interesting to consider another way to quantify the dependence of hadron masses on the
explicit breaking of chiral symmetry,  
i.e., on the pion mass squared. Let us define the quantities $A_h$ by 
\begin{equation}
\frac{\partial m_h}{\partial m_\pi^2} = \frac{A_h}{m_h}~ .
\label{eq:A}
\end{equation}
The rationale for doing this is that a parametrization of this type was used
in the past \cite{Karsch:2003zq,Karsch:2003vd,Tawfik:2005qh} taking
$A_h$ to be a constant for all hadrons heavier
than the pion
and the kaon. 
The value of this constant was estimated to be $\approx0.9-1.2$ on the basis 
of fits to data from lQCD simulations (performed at unphysical values of 
quark masses). 
One may ask
how strongly the quantities $A_h$ defined by Eq. (\ref{eq:A}) 
depend on $h$ in the model under consideration. 

The state-dependent
coefficient $A_h$ can be used to replace the sigma terms  
in the condensate formulas (\ref{eq:LightCondensate}), (\ref{eq:sCondensate})
according to (assuming $\kappa=0$)
\begin{equation}
\sigma_q^h = \frac{m_\pi^2}{m_h}A_h~,
\label{eq:sigma-A}
\end{equation}
for the light quark sigma-terms. 
Below we will also use this formula to translate sigma terms calculated
within the CQP to estimate $A_h$.
Using the GMOR relation (\ref{eq:GMORpion}), one can write for the light 
quark condensate in a HRG the compact expression
\begin{eqnarray}
\langle\bar{q}q\rangle & = & \langle\bar{q}q\rangle_0
\left(1 - \frac{A_{\rm av}n_{\rm tot}}{m_{\rm red}f_\pi^2}\right)~,
\label{eq:lightCondensate}
\end{eqnarray}
where the averaged $A_h$ coefficient is introduced as 
\begin{eqnarray}
A_{\rm av}=\frac{\sum_{h=\{M\},\{B\}} A_h n_h/m_h}
{\sum_{h=\{M\},\{B\}} n_h/m_h}~,
\label{eq:Aav}
\end{eqnarray}
while 
\begin{eqnarray}
m_{\rm red}=\left[\frac{\sum_{h=\{M\},\{B\}} n_h/m_h}{\sum_{h=\{M\},\{B\}} n_h}
\right]^{-1}
\label{mred}
\end{eqnarray}
is the weighted reduced mass and $n_{\rm tot}=\sum_{h=\{M\},\{B\}}n_h$ the 
total scalar density of hadrons.  Note, that $A_{\rm av}$ and $m_{\rm red}$
are temperature dependent.

Eq.~(\ref{eq:lightCondensate}) provides a compact expression for the 
modification of the light quark condensate in a HRG medium. 
Since $n_{\rm tot}$ and $m_{\rm red}$ are model independent 
characteristics of the HRG, the evaluation of the medium dependence 
requires solely the determination of $A_{\rm av}$ for a given model.  

\begin{figure}[!htb]
\includegraphics[width=0.45\textwidth]{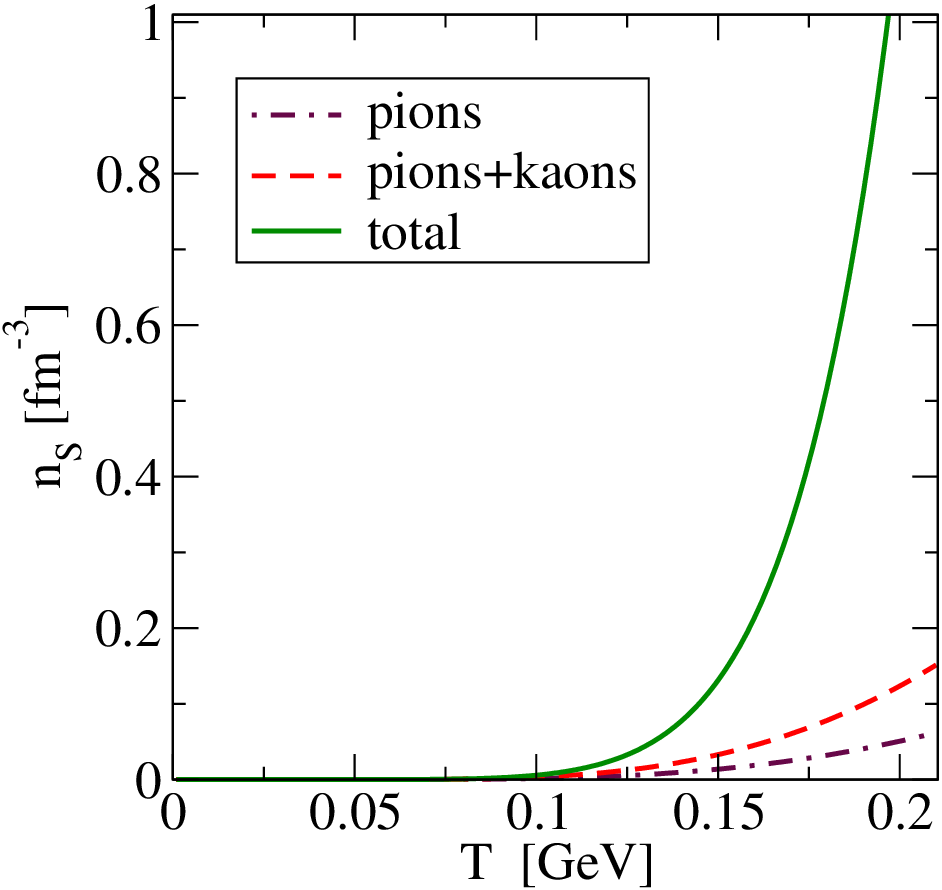}
\caption{ (Color online). Scalar densities defining the hadron contribution to
the melting of the chiral condensate.}
\label{fig:SC}
\end{figure}

The reduced mass defined in equation (\ref{mred}) is analogous 
to the reduced mass $\mu_{\rm red}$ used in many particle systems. The 
latter obeys two inequalities 
$m_{\rm lightest}/n\leq \mu_{\rm red}\leq m_{\rm lightest}$,
where $m_{\rm lightest}$ is the lightest mass in the system
of $n$ particles. Those inequalities have a direct analogy in our
case and read $m_\pi\leq m_{\rm red}\leq m_\pi n_{\rm tot}(T)/n_\pi(T)$
where the pion is the lightest hadron and $n_\pi(T)$ is the scalar density
of the pion. 
Figure ~\ref{fig:SC} shows the temperature dependence of 
the scalar densities for pions, kaons and for all hadrons included in 
the calculation.

\begin{figure}[!htb]
\includegraphics[width=0.45\textwidth]{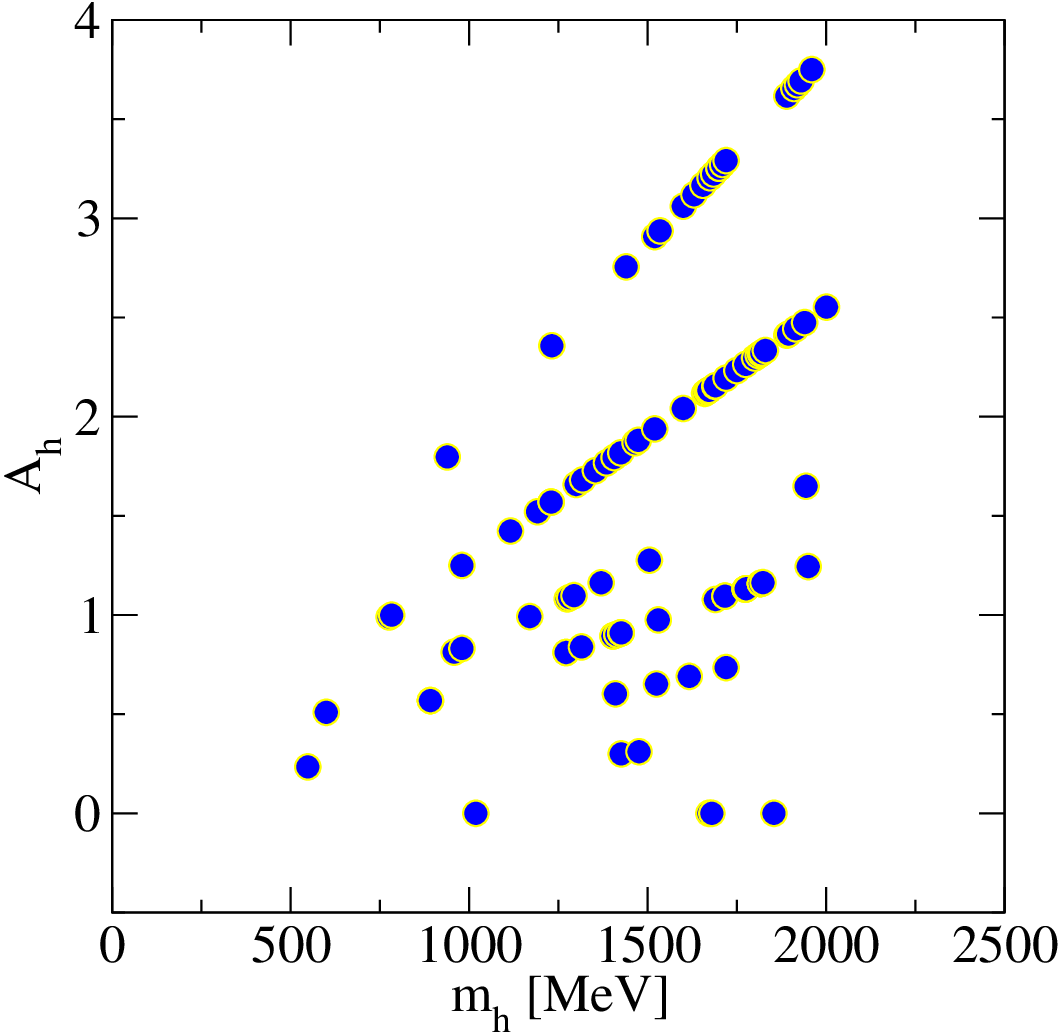}
\caption{ (Color online). Values of the $A_h$ coefficient 
for hadrons of different mass from Eq.~(\ref{eq:sigma-A}) as
evaluated with the generalized quark counting formula.}
\label{fig:A}
\end{figure}

\begin{figure*}[!htb]
\begin{minipage}{\linewidth}
\begin{minipage}{0.47\linewidth}
\includegraphics[width=0.9\textwidth]{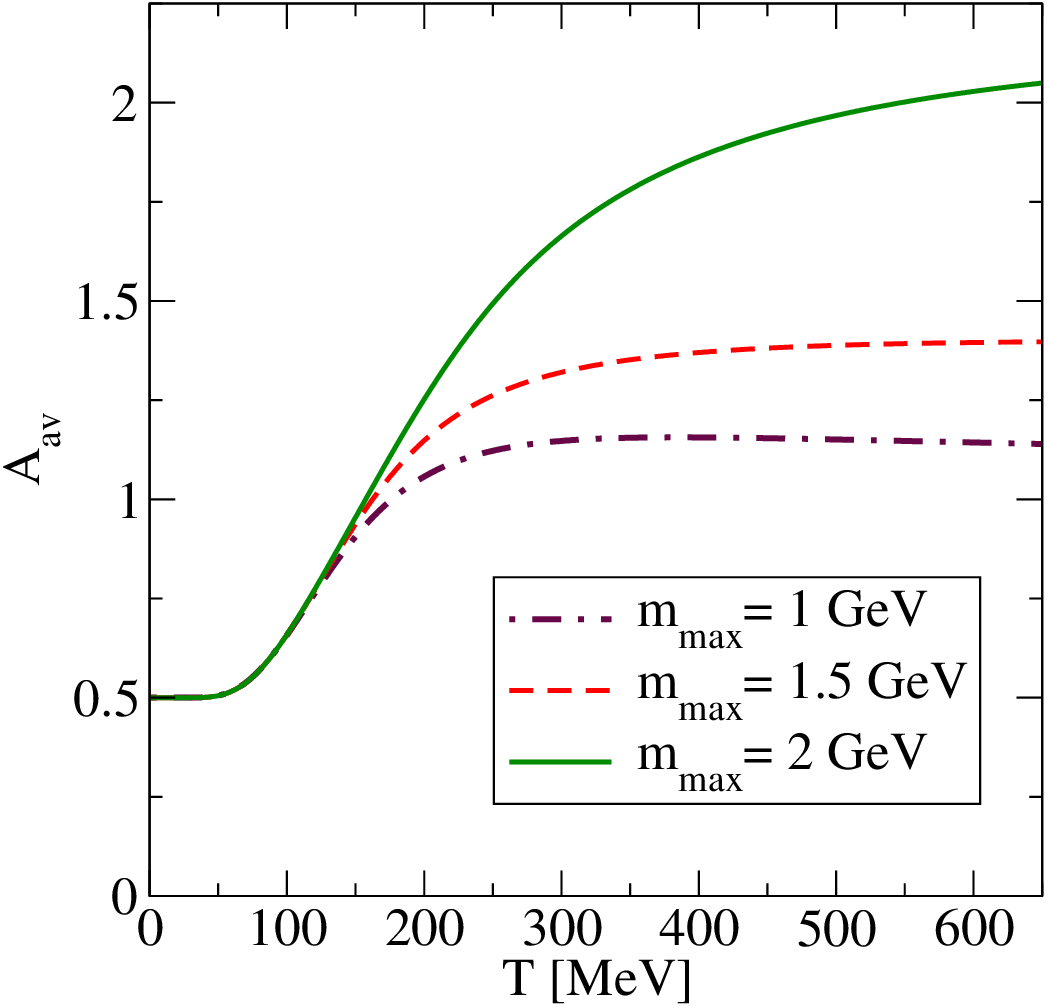}
\end{minipage}
\hspace{0.03cm}
\begin{minipage}{0.47\linewidth}
\includegraphics[width=0.9\textwidth]{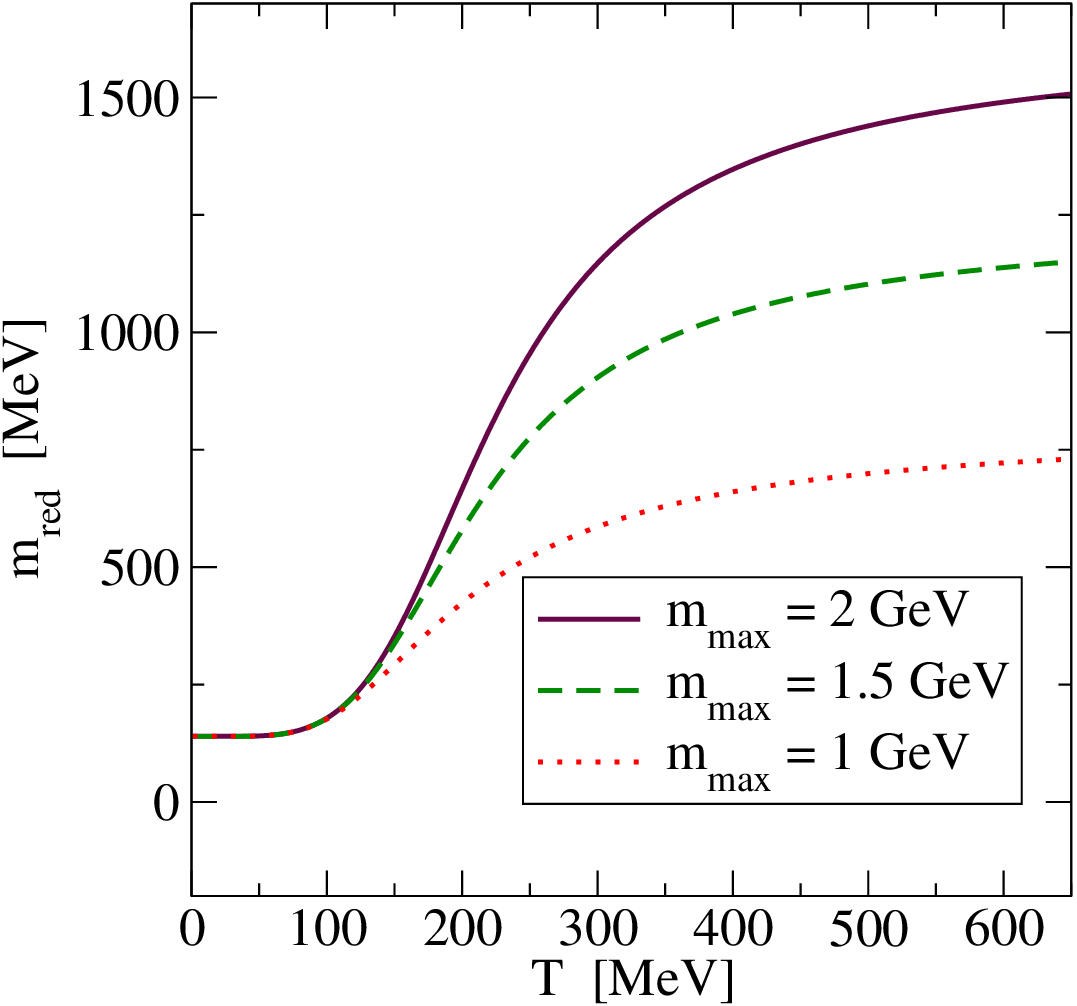} 
\end{minipage}
\end{minipage}
\caption{ (Color online). 
HRG model results. Left panel: Temperature dependence of the quantity
$A_{\rm av}$ defined by Eq.~(\ref{eq:Aav}).
Right panel: Temperature dependence of the weighted reduced mass $m_{\rm red}$ 
defined by Eq.~(\ref{mred}). 
The parameter $m_{\rm max}$ denotes the upper limit for the mass of hadrons
included in the calculation.} 
\label{fig:AL} 
\label{fig:Params}
\end{figure*}

We exemplify this for the simple quark counting model with the mass 
formulas of Eqs.~(\ref{eq:L1}) and (\ref{eq:L2}) for which we have already
given the sigma terms. The corresponding values of the $A_h$ coefficient 
as a function of hadron mass are shown in Fig.~\ref{fig:A}.
For this model, the averaged value (\ref{eq:Aav}) comes out to be 
temperature dependent and its behaviour is shown in Fig.~\ref{fig:Params}
for three different upper limits of the mass spectrum of
included hadrons.


The straight line structures of Fig. \ref{fig:A} 
reflect the fact that different hadrons admit 
different flavour structure and the assumption
that excited states have the same structure as 
their respective ground states.


Figure \ref{fig:AL} shows chiral condensates  
calculated with the two mass formulae described above. What is apparent
is that in the quark counting scenario there is a more pronounced
difference between the light and the strange condensates. In
\cite{Tawfik:2005qh} it was found that for the parametric mass 
formulae \cite{Karsch:2003vd,Karsch:2003zq} at the temperature where the light condensate 
vanishes the strange condensate is $\approx0.4$ of its vacuum value. 
The temperature where the light condensate vanishes is about
$T\approx178$~MeV.
In contrast for quark counting scheme used here this ratio is
$\langle\bar{s}s\rangle/\langle\bar{s}s\rangle_0\approx0.83$. 
The temperature where the light quark vanishes is $T\approx168$~MeV.
This difference comes from the fact that taking into account sea 
quark effects diminishes the difference between contributions from
strange and non-strange hadrons. For example nucleons would contribute
to the strange condensate and hadrons composed only of (anti)strange quarks hadrons would contribute
to the light quark condensate. This effect is captured by the parametric
mass dependence.


\begin{figure*}[!htb]
\begin{minipage}{\linewidth}
\begin{minipage}{0.47\linewidth}
\includegraphics[width=0.9\textwidth]{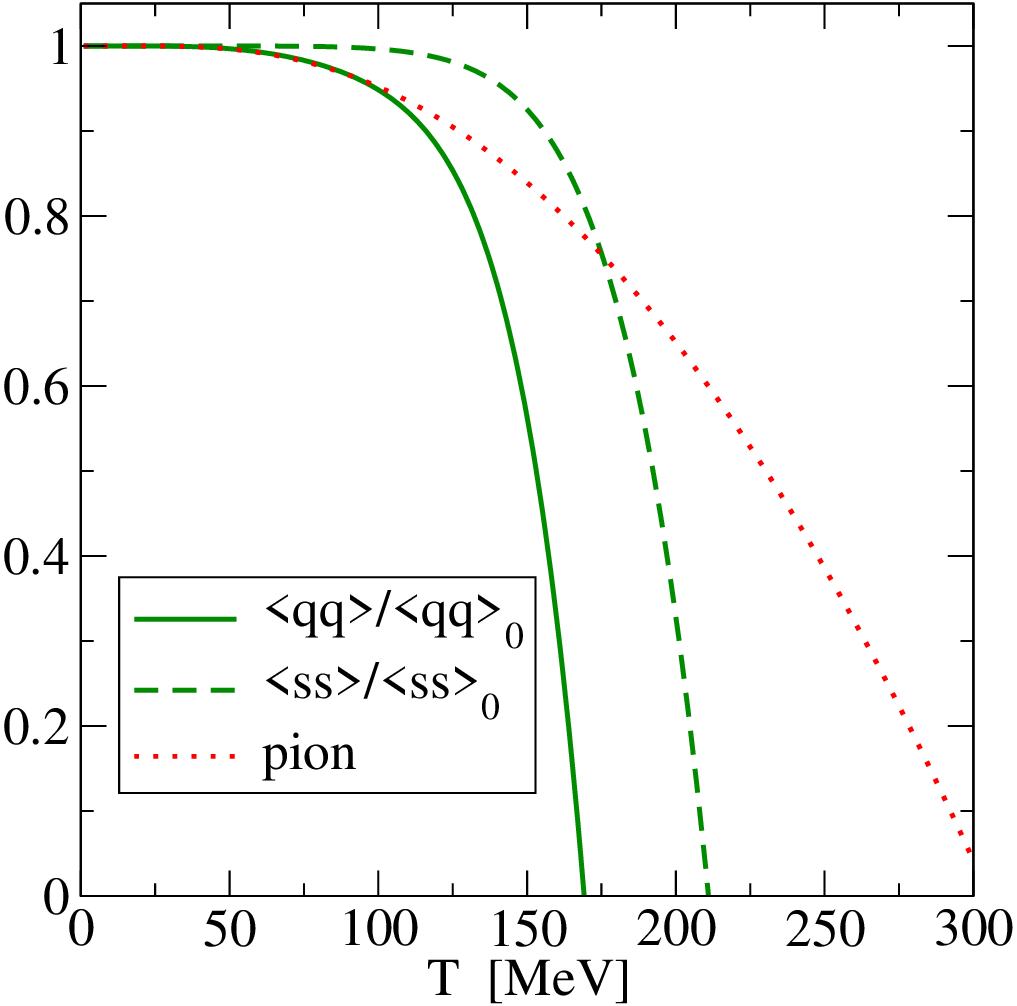}
\end{minipage}
\hspace{0.03cm}
\begin{minipage}{0.47\linewidth}
\includegraphics[width=0.9\textwidth]{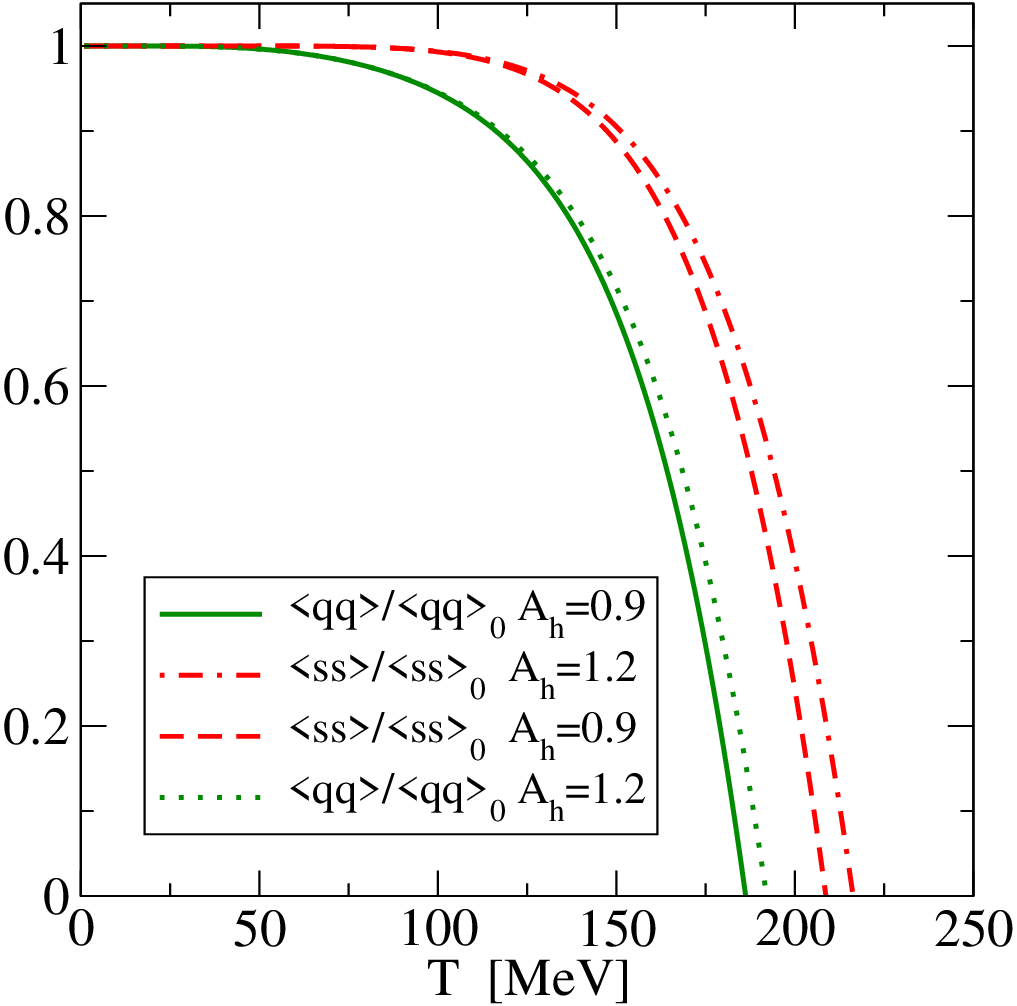} 
\end{minipage}
\end{minipage}
\caption{ (Color online). 
HRG model results. Left panel: Quark counting mass formulae based result 
for the light quark condensate (green solid line) and the strange quark 
condensate (green dashed line). 
Condensate based only on the pion gas contribution (red dotted line). 
Right panel: Chiral condensate with parametric dependence of hadron masses
\cite{Tawfik:2005qh,Karsch:2003vd,Karsch:2003zq}.} 
\label{fig:AL} 
\label{fig:HRGMcond}
\end{figure*}

\begin{figure}[!htb]
\includegraphics[width=0.45\textwidth]{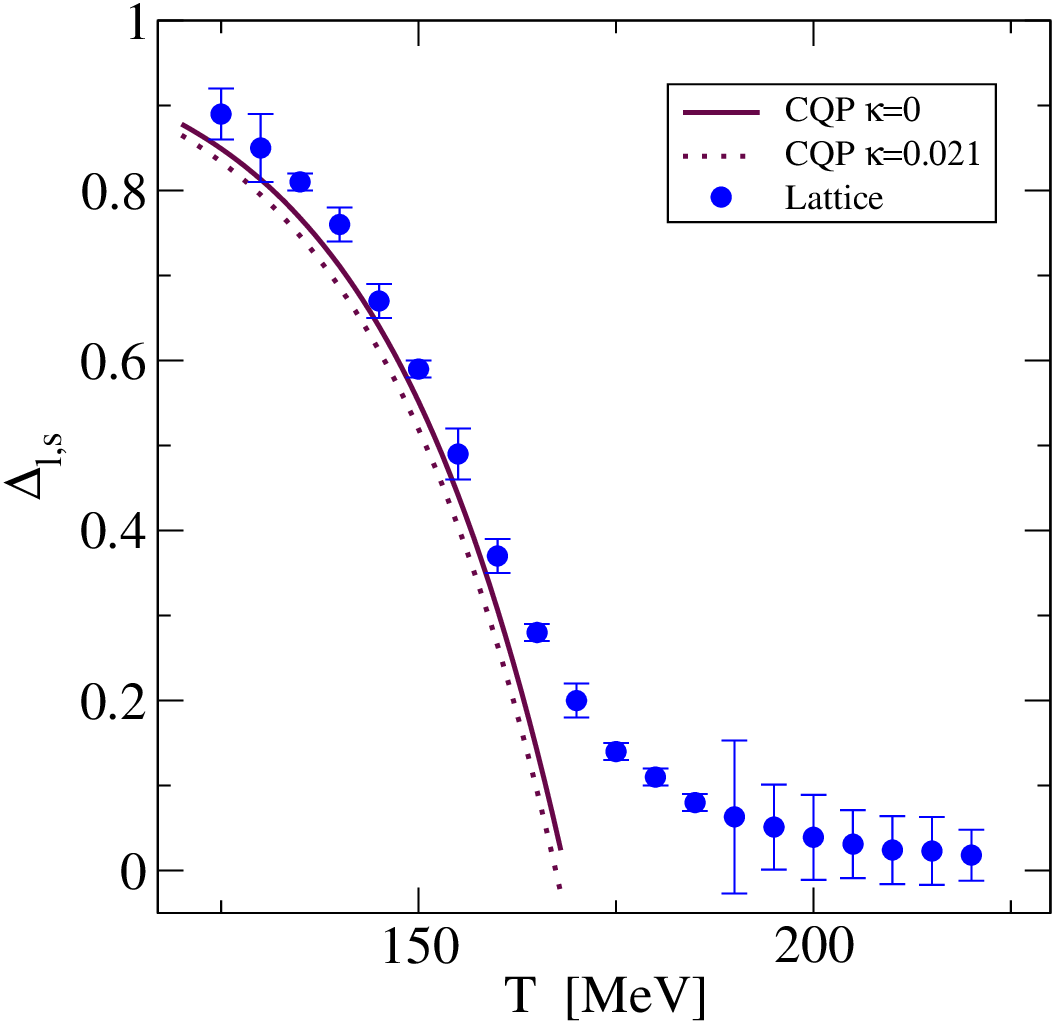}
\caption{ (Color onlline). Comparison of the HRG results for the temperature
dependence of the chiral condensate from the  constituent quark picture (CQP)
to lQCD results from the Wuppertal-Budapest collaboration 
\cite{Borsanyi:2010bp} (blue dots with error bars).}
\label{fig:DeltaK0}
\end{figure}

\begin{figure}[!htb]
\includegraphics[width=0.45\textwidth]{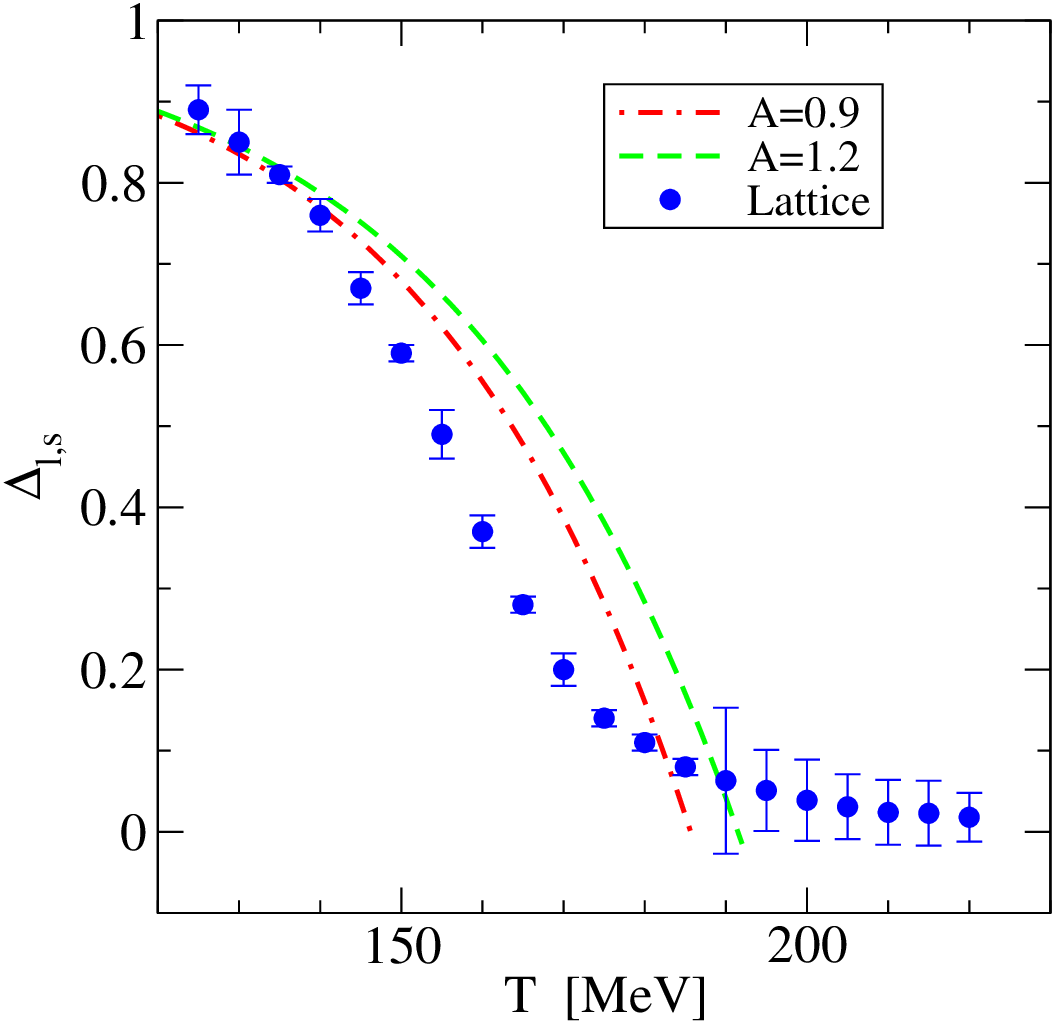}
\caption{ (Color online). 
lQCD results as in Fig.~\ref{fig:DeltaK0} (blue dots) compared to 
Comparison of the HRG results using the parametric mass formula
(red and green lines) to the lattice Wuppertal-Budapest results 
\cite{Borsanyi:2010bp} (blue dots with error bars).}
\label{fig:Delta}
\end{figure}

 
To compare with the lattice results of 
the Wuppertal-Budapest group \cite{Borsanyi:2010bp} the quantity
\begin{equation}
\Delta_{q,s}(T) = \frac{\langle\bar{q}q\rangle- \frac{m_q}{m_s}\langle\bar{s}s\rangle}{\langle\bar{q}q\rangle_0
- \frac{m_q}{m_s}\langle\bar{s}s\rangle_0}~,
\label{eq:}
\end{equation}
is considered. The reason to define this quantity on the lattice is
purely technical: in this form it eliminates a quadratic singularity at nonzero
value of the quark mass 
$m_q/a^2$ (where $a$ is the lattice spacing) and the ratio eliminates
multiplicative ambiguities  
in the definition of condensates. 
The lattice results for the $\Delta_{q,s}(T)$ are calculated for lattices with
temporal extent $N_t = 6, 10, 12$ and $16$ and an extrapolation to the 
continuum limit has been given in Ref.~\cite{Borsanyi:2010bp} to which we 
compare our models.
Physically this quantity is sensitive to chiral  
symmetry restoration: it is normalized to unity in vacuum and vanishes with
the vanishing 
of the condensates as temperature grows. 
Fig. \ref{fig:DeltaK0} shows a comparison of the lattice data 
to the HRG results with the CQP mass formulas. 
There is overall
agreement  
up to temperatures $\approx155$~MeV which is the critical temperature
from the lattice data. 
The effects of the NLO corrections on the contribution of the
pseudo-Goldstone bosons ($\kappa$ corrections) to the condensate are minor. 
To compare, in Fig. \ref{fig:Delta} the HRG results are shown together with 
those for the parametric mass formulae.
There is good agreement only for temperatures up to $\sim140$~MeV.

\section{Holographic mass formulae}
\label{Holo}

The second model considered in this paper is the holographic model of Sakai
and Sugimoto \cite{Sakai:2004cn}. This model is based on a D-brane
construction in string theory and assumes both large $N$ and large 't Hooft
coupling $g^2N$. Even though this model is neither supersymmetric nor
conformal, the approximations used are sufficiently under control to justify
the serious effort that has gone into exploring its phenomenology. Even though
in its original formulation the model did not allow for non-zero quark masses,
it leads to a large number of quantitative predictions which agree very well
with experiment despite the model having just two parameters 
\cite{Sakai:2005yt}. The
inclusion of explicit chiral symmetry breaking by non-vanishing quark masses
was studied\footnote{For an alternative approach see \cite{Bergman:2007pm}.}
in subsequent work \cite{Aharony:2008an,Hashimoto:2008sr}. 
The resulting hadron mass shifts were calculated for the 
case of two flavours in \cite{Hashimoto:2009hj} and 
for three flavours in \cite{Hashimoto:2009st}. 
The latter reference provides hadron mass formulae which were used in the 
present study. 
The results reported here include only the nucleon octet and delta decuplet 
states in the sums over hadrons,
since mass formulae have only been calculated for these states.

In the quasi-Goldstone boson sector the holographic model leads to the GMOR
formula \cite{Hashimoto:2008sr}. Although in the holographic model the 
GMOR relations were only obtained in the 
leading order $m_\pi^2=2c/f_\pi^2(m_u+m_d)$
and $m_K^2=2c/f_\pi^2(m_u+m_s)$ \cite{Hashimoto:2008sr}, in the following the 
Eqs.~(\ref{eq:GMORpion}), (\ref{eq:GMORkaon}) will be used, which include an
estimate of higher order corrections parameterized in terms of the constant
$\kappa$. 

For the baryon sector the results are as follows. In the case of two quark
flavors the 
formula  for the  nucleon octet and delta decuplet reads 
\cite{Hashimoto:2009hj}
\begin{equation}
\delta M_B = c m_\pi^2~,
\label{eq:}
\end{equation}
where $c=4.1~\textrm{GeV}^{-1}$.
The leading order
chiral perturbation theory result is
of exactly the same form with  
$c=3.6$~GeV$^{-1}$ \cite{Bernard:2003rp} 
(and references therein).
For the choice of parameters made in this paper this mass shift gives
$\delta M=80.36$~MeV and the resulting sigma term is 
\begin{equation}
\sigma= - cm_q\frac{\langle\bar{q}q\rangle_0}{f_\pi^2}
\left(1+2\kappa \frac{m_\pi^2}{f_\pi^2}\right)=36.86~\textrm{MeV}~. 
\end{equation}
Note that the above results are state independent.

The estimated pion-nucleon sigma term in chiral perturbation theory changes
from $\sigma_{\pi N}\approx59\pm19$~MeV at NLO \cite{MartinCamalich:2010fp} to
$\sigma_{\pi N}=43\pm7$~MeV at N$^3$LO, already quite close to what was
obtained above (within error bars).  There is no essential difference for this
sigma term when one includes the strange quark, which is why one can compare
this with the $2+1$ flavour results. 
In the chiral limit the nucleon octet mass was
found \cite{Borasoy:1996bx} to be $M_0=767$~MeV which gives $\delta M=171$~MeV
from the physical proton mass.

For the two-flavour DSE studies \cite{Flambaum:2005kc}
the nucleon and delta sigma terms were found to be
$\sigma_N\simeq 60$~MeV and $\sigma_\Delta\simeq 50$~MeV.
This is within the resonable limits defined by various 
model approaches but will turn out to be a little closer
to the holographic results of $2+1$ flavour case.
  
In the three flavour case \cite{Hashimoto:2009st} the nucleon octet mass
formula reads 
\begin{equation}
\delta M_N = \frac{1}{3}c_8(a_0 m_{K^0}^2 
+ a_K m_{K^\pm}^2 + a_\pi m_{\pi^\pm}^2)~,
\label{eq:Mass8}
\end{equation}
and for the delta decuplet
\begin{equation}
\delta M_\Delta = \frac{1}{3}c_{10}(a_0 m_{K^0}^2 
+ a_K m_{K^\pm}^2 + a_\pi m_{\pi^\pm}^2)~,
\label{eq:Mass10}
\end{equation}
where $c_8=7.9~\textrm{GeV}^{-1}$, $c_{10}=9.5~\textrm{GeV}^{-1}$ and 
the $a$ coefficients are given in tables \ref{Tab:aN} and \ref{Tab:aDelta}. 
Using equations (\ref{eq:PionDer}) and (\ref{eq:KaonDer}) one can calculate 
derivatives of baryon masses  
\begin{eqnarray}
\frac{\partial(\delta M_B)}{\partial m_u} & =& \nonumber
\frac{1}{3}c_\#\Bigg[a_K\frac{\langle\bar{q}q\rangle_0 +
    \langle\bar{s}s\rangle_0}{2f_K^2}\left(1+2\kappa\frac{m_K^2}{f_\pi^2}\right)\\  
&+& a_\pi\frac{\langle\bar{q}q\rangle_0}{2f_\pi^2}\left(1+2\kappa\frac{m_\pi^2}{f_\pi^2}\right)
\Bigg]~,
\label{eq:}
\end{eqnarray}
\begin{eqnarray}
\frac{\partial (\delta M_B)}{\partial m_d} &= & \nonumber \frac{1}{3}c_\#\Bigg[a_0\frac{\langle\bar{q}q\rangle_0+\langle\bar{s}s\rangle_0}{2f_K^2}\left(1+2\kappa\frac{m_K^2}{f_\pi^2}\right)\\
&+&
a_\pi\frac{\langle\bar{q}q\rangle_0}{2f_\pi^2}\left(1+2\kappa\frac{m_\pi^2}{f_\pi^2}\right)
\Bigg]~,
\label{eq:}
\end{eqnarray}
\begin{equation}
\frac{\partial (\delta M_B)}{\partial m_s} = \frac{1}{3}c_\#(a_0+a_K)\frac{\langle\bar{q}q\rangle_0+\langle\bar{s}s\rangle_0}{2f_K^2}\left(1+2\kappa\frac{m_K^2}{f_\pi^2}\right)~,
\label{eq:}
\end{equation}
where $B=N,\Delta$ and $\#=8,10$. 



\begin{table}[h]
	\centering
		\begin{tabular}{c|*{8}{c}}
			 {\bf 8}       & P        &    N       &  $\Lambda$   &   $\Sigma^+$ &  $\Sigma^0$ &  $\Sigma^-$ & $\Xi^0$  & $\Xi^-$    \\ \hline 
			 $a_0$    &    3/5        &    4/5     &    9/10      &        3/5   &      11/10  &     8/5     &   4/5    &   8/5    \\ 
			 $a_K$    &    4/5        &    3/5     &    9/10      &       8/5    &  11/10      &     3/5     &    8/5   &  4/5   \\ 
			 $a_\pi$  &  8/5          &    8/5     &    6/5       &       4/5    &    4/5      &     4/5     &     3/5  & 3/5
			 \end{tabular}
		\caption{Coefficients in the nucleon mass formula
			\label{Tab:aN}}
\end{table}

\begin{table}
	\centering
		\begin{tabular}{c|*{10}{c}}
{\bf 10} & $\Delta^{++}$  &  $\Delta^+$ &  $\Delta^0$ &   $\Delta^-$ &  $\Sigma^{*+}$ & $\Sigma^{*0}$ & $\Sigma^{*-}$ & $\Xi^{*0}$  & $\Xi^{*-}$ & $\Omega^-$      \\ \hline
			 $a_0$    & 1/2 & 3/4 & 1 & 5/4 & 3/4 & 1 & 5/4 & 1 & 5/4 & 5/4\\ 
			 $a_K$    &5/4 & 1 & 3/4 & 1/2 & 5/4 & 1 & 3/4 & 5/4 & 1 & 5/4 \\ 
			 $a_\pi$    & 5/4 & 5/4 & 5/4 & 5/4 & 1 & 1 & 1 & 3/4 & 3/4 & 1/2
		\end{tabular}
		\caption{Coefficients in the delta mass term
			\label{Tab:aDelta}}
\end{table}


The resulting hadronic sigma terms are presented in the tables
\ref{tab:sigma1},\ref{tab:sigma2}.

In the holographic setup the strange nucleon sigma term is significantly
overestimated, indicating that higher order corrections are needed. 
For ChPT the leading order tree level result is expressed in terms of five 
low-energy constants and gives a reasonably good evaluation of the nucleon 
strange sigma term. 
For the purpose of comparison, the results for the leading order ChPT
sigma terms are shown in table \ref{tab:sigmaChPT}. 
The strange sigma term for the nucleon, $\sigma_s^N\approx 162$~MeV, is a bit 
large but still reasonable.  
It should be noted that when compared to the NLO results from
\cite{MartinCamalich:2010fp} even the sign of the sigma terms can change,
meaning that higher order corrections cannot be ignored. 
It is to be expected that including higher order corrections in the 
holographic approach should cure the problem of overestimating the strangeness 
contribution as it does in the case of ChPT. 

\begin{center}
\begin{figure}[!htb]
\begin{minipage}{\linewidth}
\begin{minipage}{0.47\linewidth}
\includegraphics[width=.9\textwidth]{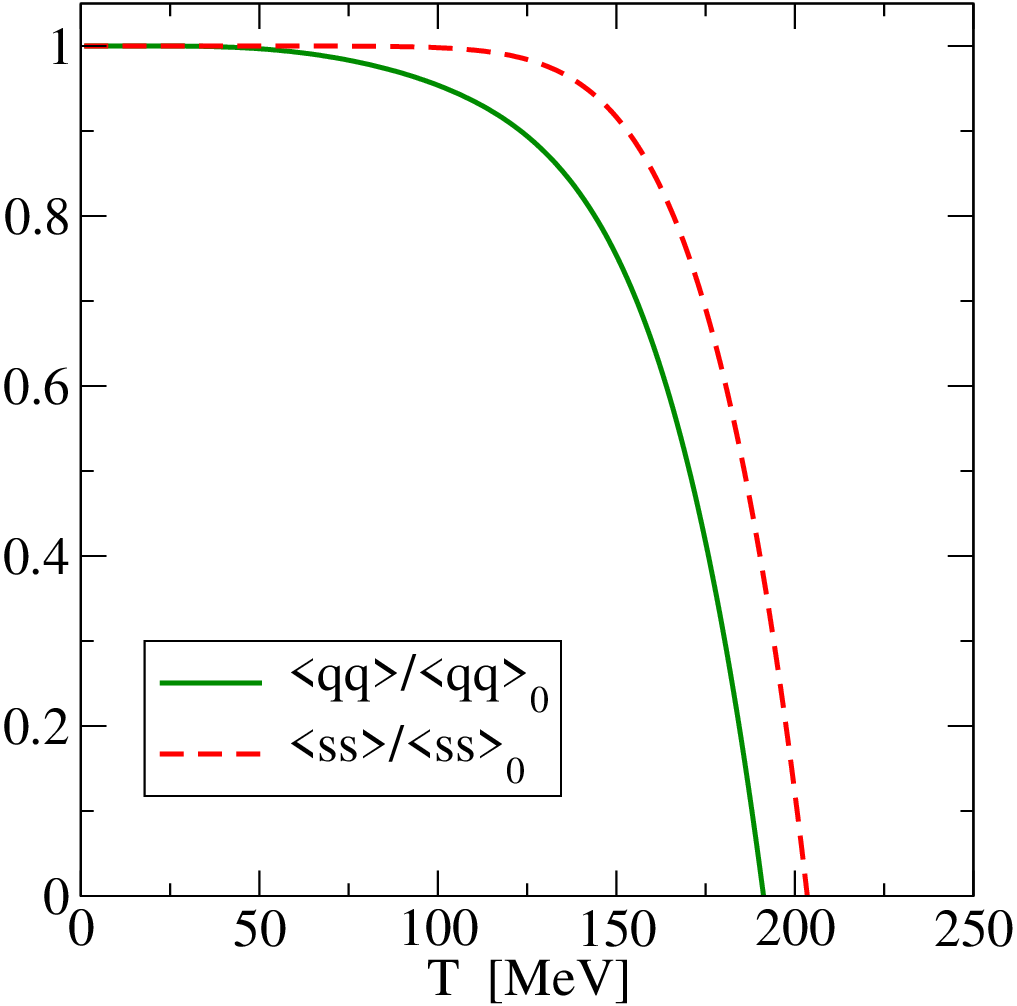}
\end{minipage}
\hspace{0.03cm}
\begin{minipage}{0.47\linewidth}
\includegraphics[width=0.9\textwidth]{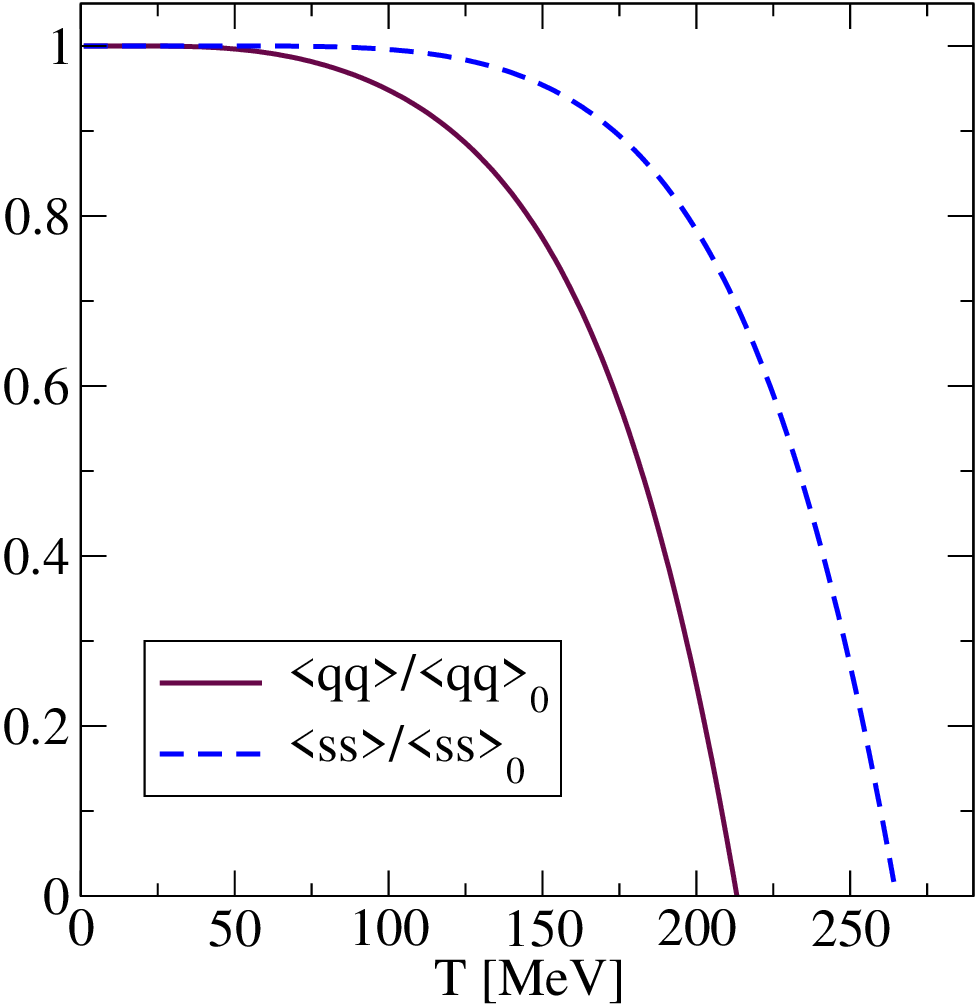} 
\end{minipage}
\end{minipage}
\caption{ (Color online). Temperature dependence of strange and light condensates 
using holographic mass formulae (left panel) and  
NLO ChPT mass formulae  \cite{MartinCamalich:2010fp} (right panel).
For details, see the text. }
\label{fig:LBCond}
\end{figure}
\end{center}

Fig.~\ref{fig:LBCond} presents the result for the chiral condensates
obtained with holographic and NLO ChPT mass formulae where only the nucleon
octet and delta decuplet baryons are included (apart from the quasi-Goldstone
bosons).  
In the holographic case, due to the overestimated sigma terms, the
difference between strange and light condensates is diminished. 
For the same reason the too small number of states included in the strange 
sector is compensated.

\onecolumngrid
\begin{center}
\begin{table}[!htb]
	\centering
		\begin{tabular}{c|*{8}{c}}
			 {\bf 8}       & P        &    N       &  $\Lambda$   &   $\Sigma^+$ &  $\Sigma^0$ &  $\Sigma^-$ & $\Xi^0$  & $\Xi^-$    \\ \hline 
			 $\sigma_u$    & 50.5611  &   47.3916  &   42.6751    &   44.2976    &  36.3738    &   28.4501   & 39.5622  & 26.8842     \\ 
			 $\sigma_d$    & 47.3916  &   50.5611  &   42.6751    &   28.4501    &  36.3738    &   44.2976   & 26.8842  & 39.5622   \\ 
			 $\sigma_s$    & 1189.1   &   1268.62  &   1070.76    &   713.838    &  912.652    &   1111.47   & 674.548  & 992.651 
		\end{tabular}
		\caption{Sigma terms for the nucleon octet in MeV.}
		\label{tab:sigma1}
\end{table}
\end{center}
\twocolumngrid


\onecolumngrid
\begin{center}
\begin{table}[h!]
		\begin{tabular}{c|*{10}{c}}
{\bf 10} & $\Delta^{++}$  &  $\Delta^+$ &  $\Delta^0$ &   $\Delta^-$ &  $\Sigma^{*+}$ & $\Sigma^{*0}$ & $\Sigma^{*-}$ & $\Xi^{*0}$  & $\Xi^{*-}$ & $\Omega^-$      \\ \hline
			 $\sigma_u$    & 49.4056 & 45.4437 & 41.4818 & 37.5199 & 43.4863 & 39.5245 & 35.5626 &
37.5671 & 33.6052 & 31.6479 \\ 
			 $\sigma_d$    & 37.5199 & 41.4818 & 45.4437 & 49.4056 & 35.5626 & 39.5245 & 43.4863 &
33.6052 & 37.5671 & 31.6479\\ 
			 $\sigma_s$    & 941.409 & 1040.82 & 1140.22 & 1239.63 & 892.297 & 991.705 & 1091.11 &
843.186 & 942.593 & 794.074
		\end{tabular}
		\caption{Sigma terms for the delta decuplet in MeV.}
		\label{tab:sigma2}
\end{table}
\end{center}
\twocolumngrid


\onecolumngrid
\twocolumngrid


As is clear from the above discussion, one important extension of the existing
calculations in the holographic model would be to calculate higher order 
corrections in the current quark masses. 
One motivation for it was already mentioned: this would improve the 
resulting sigma terms, especially in the strange sector. 
A second, more formal, motivation is that in ChPT for $N_f=2$ the
second order correction to the proton mass has the universal form 
\cite{Bernard:2003rp}
\begin{equation}
M_N^{(3)} = M_0 + 4c_1m_\pi^2 + \frac{3g_A^3}{32\pi f_\pi^2}m_\pi^3~,
\label{eq:}
\end{equation}
where $g_A$ is the axial coupling. 
If one adopts the usual scaling of parameters with $N$, then one 
gets $g_A\sim N$ and $f_\pi^2\sim N$, so that the subleading contribution
would scale like $\sim N^2$ which would dominate the leading order result 
$M_0\sim N$. 
On the other hand, if one follows recent argumentation \cite{Hidaka:2010ph}
that $g_A\sim N^0=1$ then NLO contributions would be of the order $\sim 1/N$.
It would be interesting to check if in the Sakai-Sugimoto model this 
universality also holds.


\section{Conclusions}
\label{Conclusions}

This paper was devoted to a discussion of the finite temperature behaviour 
of the chiral condensate within the HRG framework exploring different 
microscopic descriptions of the dependence of hadron masses 
on the current quark mass. 
In particular, a constituent quark scheme and holographic mass formulae have 
been used. 
It was also studied how the results are affected by including different 
numbers of states in the sums over resonances. 
It turns out that with a sensible choice of mass formulae and including 
hadron states with masses up to $\sim2$~GeV generic agreement
with recent lattice results is obtained. 
This is yet another confirmation of the well known fact that for low 
temperatures the HRG model gives a satisfactory physical interpretation of 
lQCD data.
Chiral symmetry restoration in the strange sector was seen to take place at 
higher temperatures than in the light quark sector \cite{Kunihiro:1987bb,Kogut:1988zc}, 
which is related
both to the lower number of strange hadrons contributing to the condensate as 
well as to the response of hadron masses to changes in the current strange 
quark mass. 

A generalization of the quark-counting approach of 
\cite{Blaschke:2011hm,Leupold:2006ih} was proposed, and it was shown that 
the mass relations where only valence quarks of the hadron are taken into 
account already lead to a behaviour of the condensate
which is close to what is seen in the full lattice data. 
In this scheme dynamically generated (constituent) quark masses
are considered and their dependence on the current quark mass is quantified 
in the framework of the NJL model. 
This step takes into account part of the non-perturbative QCD dynamics.  
The sea quark contributions are neglected resulting in a vanishing strange 
sigma term for the nucleon and a vanishing light quark contribution for the
$\Omega^-$ baryon. 
This is somewhat in the spirit of the large-$N$ expansion where quark loops 
are suppressed.

Along with this, a careful analysis of the hidden strange mesons has been 
performed based on the flavour symmetry structure of the mesons. 
This affects the simple quark counting rules used by 
\cite{Blaschke:2011hm,Leupold:2006ih},
taking into account neglected effects which overestimated the light quark 
condensate and underestimated the strange quark condensate.

Another new aspect considered in this paper concerns the sigma terms and the
condensate following from the mass formulae of the holographic model of QCD 
due to Sakai and Sugimoto \cite{Sakai:2004cn}. 
These formulae take on a form similar to the tree level ChPT results with 
strange sigma terms overestimated due to the inaccuracy of the
approximation for the relatively large value of $m_s$. 
Since those shifts were only calculated for the nucleon octet and delta 
decuplet baryons, the computation of the condensate is incomplete. 
This also shows the importance of heavier hadrons for temperatures near the 
QCD transition temperature.
  
The results obtained here are of great importance in the context  
of hadron production under extreme conditions in heavy-ion collisions.
Recently, it has been conjectured that the behaviour of the chiral condensate  
determines the collision rates of hadrons and thus may provide a microscopic 
approach to the chemical freeze-out of hadron species
\cite{Blaschke:2011hm}. 
This approach, however, has yet been considered only in the light quark 
sector.
Including the strange quark condensate in that analysis could advance the 
understanding of strangeness production in heavy ion collision experiments.


\acknowledgments
We thank M. Gazdzicki, M. P. Heller and C. D. Roberts for useful 
discussions. 
We are grateful to N. Agasian, J. M. Alarcon, K. Hashimoto, O. Louren\c{c}o,
E. Megias, and J. R. Pelaez for their comments to this paper. 
D.B. is grateful for the hospitality and support during his stay at the 
University of Bielefeld.
He acknowledges partial support from the Narodowe Centrum Nauki
(NCN) within the "Maestro" programme under contract No. 
DEC-2011/02/A/ST2/00306 and from the Russian Foundation
for Basic Research (RFBR) under grant No. 11-02-01538a. 
J.J. received support from the Polish Ministry of Science and
Higher Education under grant No. 8975/E-344/M/2012 for young scientists.

\newpage
\thispagestyle{empty}




\end{document}